\documentclass{osa-article}

\journal{osajournal}

\articletype{Research Article}
\usepackage{parskip}
\usepackage{todonotes, subcaption, rotating}

\graphicspath{{../../Plots/}}

\begin{document}

\title{InSPECtor: an end-to-end design framework for compressive pixelated hyperspectral instruments}

\author{T.A. Stockmans\authormark{1, *}, F. Snik\authormark{1}, M. Esposito \authormark{2},
C. van Dijk \authormark{2}, C.U. Keller\authormark{1, 3}}

\address{\authormark{1}Leiden Observatory, Leiden University,  P.O. Box 9513, 2300 RA Leiden, The Netherlands\\
\authormark{2}cosine Remote Sensing, Warmonderweg 14, 2171 AH  Sassenheim, The Netherlands\\
\authormark{3}Lowell Observatory, 1400 W Mars Hill Rd, Flagstaff, AZ 86001, USA}

\email{\authormark{*}stockmans@strw.leidenuniv.nl} 



\begin{abstract}
Classic designs of hyperspectral instrumentation densely sample the spatial and spectral information of the scene of interest. Data may be compressed after the acquisition.
In this paper we introduce a framework for the design of an optimized, micro-patterned snapshot hyperspectral imager that acquires an optimized subset of the spatial and spectral information in the scene. The data is thereby compressed already at the sensor level, but can be restored to the full hyperspectral data cube by the jointly optimized reconstructor. 
This framework is implemented with TensorFlow and makes use of its automatic differentiation for the joint optimization of the layout of the micro-patterned filter array as well as the reconstructor. We explore the achievable compression ratio for different numbers of filter passbands, number of scanning frames, and filter layouts using data collected by the Hyperscout instrument.
We show resulting instrument designs that take snapshot measurements without losing significant information while  reducing the data volume, acquisition time, or detector space by a factor of 40 as compared to classic, dense sampling. 
The joint optimization of a compressive hyperspectral imager design and the accompanying reconstructor provides an avenue to substantially reduce the data volume from hyperspectral imagers. 
\end{abstract}

\section{Introduction}
\label{chapter: Introduction}

Hyperspectral imaging combines the acquisition of two-dimensional spatial and spectral information; \cite{bioucas-dias_hyperspectral_2012} it is used in a broad range of research, including -- but not limited to -- remote sensing \cite{vane_airborne_1993,barnes_status_2003}, food quality control \cite{elmasry_chapter_2010,gowen_hyperspectral_2007}, archaeology \cite{liang_advances_2012}, astronomy \cite{hege_hyperspectral_2004}, agriculture \cite{adao_hyperspectral_2017}, medical imaging \cite{fei_chapter_2020,ortega_hyperspectral_2020}, and imaging on micro-scales for biological and chemical processes \cite{dong_dmd-based_2021}.

Hyperspectral imaging contains three dimensions of information (two spatial and one spectral dimension), but most detectors are two-dimensional. This requires a trade-off in the instrument design, which often results in using time as the third dimension. The three most common techniques for air- and space-borne instruments are whisk broom, push broom (line-scan), and staring \cite{eismann_hyperspectral_2012}. In the whisk broom imaging mode, the system measures the full spectrum of one geometrical pixel before stepping to the next in a track perpendicular to the flight direction. In the push broom mode, the system simultaneously measures the spectrum of a line of geometrical pixels. Staring, as opposed to the other two modes, measures the whole image in one spectral band and then steps through the bands \cite{li_review_2013, eismann_hyperspectral_2012, ustin_current_2021}.

The techniques described above all have in common that the acquired hyperspectral data cube is densely sampled and therefore partially redundant \cite{willett_sparsity_2014}. This redundancy implies that there is a representation of the data cube in which most entries map to (approximately) zero and can be ignored. When only measuring the non-zero entries of this representation, all information could still be recovered while reducing detector space, data volume, and/or measurement time. Such reductions are particularly helpful for applications in space \cite{guzzi_donatella_optical_2019}, where mass, volume, power, and data rates are limited. A simple example of the redundancy in hyperspectral data cubes is the success of hyperspectral band selection where the algorithms extract the spectral bands that contain the most information \cite{sun_hyperspectral_2019}. However, this is a post-processing method that does not improve the detector size and/or the acquisition time.

The imaging techniques that go beyond dense sampling are typically referred to as \textit{compressed sensing (CS)}. Several hyperspectral instruments based on CS have been designed \cite{cao_computational_2016}. Examples include the space-borne concepts proposed by \cite{coluccia_optical_2020, barducci_compressive_2014} and the \textit{computed tomographic imaging spectrometry (CTIS)} system \cite{hege_hyperspectral_2004} based on \cite{okamoto_simultaneous_1991}. The most common compressive sensor for hyperspectral imaging is the \textit{coded aperture snapshot spectral imager (CASSI)} and its variations \cite{wagadarikar_single_2008, arce_compressive_2014, gehm_single-shot_2007, wu_development_2011,august_compressive_2013}, which combine spectral dispersers and coded focal-plane masks. Other designs combine coded focal-plane masks with a dispersing lens \cite{kar_compressive_2019}, a diffuser in combination with a \textit{color filter array (CFA)} \cite{monakhova_spectral_2020}, or a Fourier Transform Spectrometer and a single-pixel detector \cite{jin_hyperspectral_2017}.

A compressive sensor can be described mathematically by a single matrix $\bf{H}$. This measurement matrix, when multiplied with the vector representation of the measured scene, $\vec{x}$, and adding the noise, $\vec{n}$, results in the vector representation of the detected signal, $\vec{y}$, i.e.\ 

\begin{equation}
	\label{equation: CS}
	\vec{y}= \mathbf{H} \vec{x}+\vec{n}\;.
\end{equation} \noindent

The reconstructor estimates $\vec{x}$ from knowing $\vec{y}$ and $\bf{H}$. This inversion is not trivial due to the non-uniqueness of the problem, which requires the addition of constraints. Examples of such reconstruction algorithms can be found in \cite{tropp_computational_2010, wang_compressed_2015, yang_hyperspectral_2021} and references therein. 

Some authors have optimized the instrument and thereby the measurement matrix; they require the most accurate reconstructions, whilst operating the fastest or sparsest \cite{gozcu_learning-based_2018,wu_learning_2019,li_learning_2016, baldassarre_learning-based_2016}. The numerical optimization of the imaging system, in light of the needed reconstruction, resembles the definition of \textit{Computational Imaging (CI)}. This should not be surprising since the field of CS has been closely intertwined with the field of CI for some time now \cite{mait_computational_2018}. CI has also produced efficient and relevant instruments, like the one described in \cite{gao_computational_2022} or in \cite{arguello_deep_2022}. An overview of some instruments in this field that use deep learning mostly for the reconstruction of the hyperspectral data cube can be found in \cite{huang_spectral_2022} and \cite{bacca_computational_2023}. Finally, there is the work of Wang et al. \cite{wang_hyperreconnet_2019} where the reconstruction was jointly optimized with the coded aperture mask of the CASSI system.

Another research field that is related to CS is called demosaicing \cite{kaur_survey_2015}. It can be described within the compressed-sensing framework as a specific type of reconstructor, but this is not normally done due to its origin and scope in \textit{Red-Green-Blue (RGB)} photography. In most common RGB instruments, the detector is covered by a Bayer pattern of filters \cite{bayer_color_1976}, i.e.\  red, green, and blue pixels are alternating in a 2x2 super-pixel, where the green filter occurs twice. When a scene is imaged with a single snapshot, the intensity of the blue and green light at the location of a red pixel is unknown, and vice versa. Demosaicing provides an approximation of the intensities of all the unknown colors of each pixel, creating a fully filled RGB image. Some examples of demosaicing algorithms for RGB imaging can be found in \cite{gharbi_deep_2016,cui_color_2018,guo_joint_2021,he_self-learning_2012,heinze_joint_2012,iriyama_deep_2021,jin_review_2020,kaur_survey_2015,kiku_beyond_2016,menon_color_2011,sharif_beyond_2021,wang_multilayer_2014,wu_color_2011,zhang_learning_2018}. Since its beginning in RGB imaging, demosaicing has also found its way to detectors with more spectral bands than only the RGB broadband filters. Examples of demosaicing algorithms dealing with multispectral data cubes can be found in  \cite{habtegebrial_deep_2019,dijkstra_hyperspectral_2019,li_deep_2021,wang_discrete_2013, zhuang_hy-demosaicing_2022, tsagkatakis_graph_2019, mihoubi_multispectral_2017,amba_n-lmmse_2017, arad_ntire_2022}. 

The demosaicing algorithms we referenced above make use of existing color filter array designs and instruments. The design of these \textit{color filter arrays (CFAs)} themselves has also undergone development. The CFA design can be updated to enhance the quality of the resulting processed images. For instance, in the update of the Bayer layout for RGB imagers, we refer to \cite{lukac_color_2005, hirakawa_spatio-spectral_2008, li_optimized_2017} and other references in the latter publication. Some examples of multi-spectral CFA designs can be found in \cite{miao_design_2006, li_optimized_2018,lapray_multispectral_2014, saxe_advances_2018}. Finally, some of the commercially available instruments for snapshot hyperspectral imaging with a CFA include the SNAPCAM by \cite{pichette_fast_2017}, XIMEA's detector \cite{lemmens_combination_2020}, and Silios' detector \cite{cheng_-line_2018}. An older, but more general overview of snapshot spectral imagers is given in \cite{hagen_review_2013}.

Some authors have also directly related RGB imaging and hyperspectral imaging, which is referred to as spectral recovery. In this field, RGB images are transformed into hyperspectral data cubes \cite{fu_hyperspectral_2020, arad_ntire_2022-1,tao_compressive_2021}.

In the papers referenced above, the measurement design (instrument) and the reconstruction algorithm are treated as separate entities. By considering the system as a whole and jointly optimizing the instrument design and the reconstruction algorithm together, advances have been made in RGB imaging \cite{chakrabarti_learning_2016,henz_deep_2018}. The joint optimization of instrument design and recovering algorithm is also done by \cite{jacome_d2uf_2022}. The instrument they have optimized combines the CASSI system described above and a multi-spectral filter array. The broadband encoding stochastic (BEST) camera descibed in \cite{zhang_deeply_2021} and \cite{song_deep-learned_2021} is developed by a neural network which designs both the spectral filters of a hyperspectral camera and the dense neural network for the reconstruction afterwards. Finally, most closely related to this work is the conference paper by Li, Dai and Van Gool \cite{li_jointly_2023}.  They describe the use of a reinforcement learning based band selection algorithm in combination with a neural network to design the hyperspectral CFA and do the reconstruction afterwards. There are 3 key points that make the work described here different: 1) They do a band selection of common broadband filters. In our described framework, the spectral properties of the filters can be set as another optimizable parameter, which enables a bigger versatility. 2) Their reconstruction network first demosaics the images from the CFA and then uses a separate spectral recovery algorithm to reconstruct the hyperspectral datacube. We combine this step in a single mapping, which reduces the chance of error propagation between separate layers. 3) Finally, they solely focus on snapshot imaging. The methods we describe below also take the possibility for push broom scanning into account, which can push the accuracy of the instrument for some applications far enough to make it a feasible alternative to classical hyperspectral instruments. 

In this paper, we describe a new framework that can jointly design a spectral filter array and reconstruction function that combine into a compressive hyperspectral imager. The papers mentioned above either optimize only one part of these two or are focussed on a different optical design altogether. The presented framework can optimize three aspects: 1) it optimizes the filters that contain the most spectral information; 2) it determines the layout of these filters to optimize the estimated spectra for all pixels; and 3) it determines a linear reconstructor that optimally demosaics the measurements into a filled hyperspectral data cube. In the following chapter, we describe the framework in detail. Then we show the accuracy provided by designs that vary in terms of the number of filters and the number of push broom scans. Finally, we present an outlook for future applications and improvements. 

\section{Methods}
To design the optimal compressive measurement set-up and reconstructor, we developed the InSPECtor framework. Currently, InSPECtor consists of two components. The first component only takes the spectral dimension into account and disregards the spatial dimension of a hyperspectral dataset. This component determines a given number of optimized spectral filter passbands to reconstruct the full spectra. The second component, however, takes both the spatial and spectral dimensions into account. It can, for instance, decide on the optimal layout of the filters from the first component and return the matching reconstructor. The second component can also be used to decide the best passbands in a specific fixed layout and optimize a reconstructor for the resulting instrument. In the future, the optimization of both the filters and their layout will be combined in a single framework. Below we start with an explanation of the merit functions used in our paper. We continue with a mathematical formulation of the two components in a compressed sensing formulated manner. The end of this chapter entails the implementation of the two components in Python code using the TensorFlow package.

\subsection{Merit functions}
To determine the accuracy of the resulting data cubes as compared to the original ones, we calculate the Mean Square Error (MSE) and the Peak Signal to Noise Ratio (PSNR). The PSNR is a widely used metric for spectral image comparisons\cite{hore_image_2010}. We included the MSE to provide a non-logarithmic scale of the error for direct comparison to the spectra. The MSE is defined as follows:

\begin{equation}
	MSE = \frac{1}{M}\sum_{k = 0}^M (Y_k-P_k)^2\;,
\end{equation} \noindent

in which $Y$ is the true scene and $P$ is the estimated scene and $M$ is the total number of entries that the scene consists of. This scene can be either a single spectrum or the spectra of multiple spatial pixels. The PSNR is closely related to the MSE in a logarithmic inverse way. So note that a lower MSE means a better estimation and corresponds to a higher PSNR. The mathematical formulation of the PSNR is as follows:

\begin{equation}
	PSNR = 10 \log_{10}(\frac{MAX^2}{MSE})\;,
\end{equation} \noindent

where $MAX$ is the maximum possible value of $Y$. $MAX$ differs between applications and scenes under investigation. For example, for an 8-bit system, $MAX=255$. In this paper, we have used $MAX=2^{32}-1$ since the images in the training and test data sets are 32-bit. 

\subsection{Optimal filters estimator}
\label{section: optimal filters}
The first component of InSPECtor determines the optimal filter passbands independent of the spatial arrangement of the filters. The filters are implemented as a linear transformation from an input spectrum to filtered intensity measurements, from which a linear reconstructor estimates the input spectrum. 

The linear transformation from an input spectrum to filtered intensity measurements is mathematically equivalent to equation \ref{equation: CS}. Here, $\vec{x}$ is the discretely sampled spectrum, $\vec{y}$ are the intensities through every filter, and $\bf{H}$ can be described as:

\begin{align}
	\label{equation: best filters matrix}
	\bf{H} &= 
	\begin{pmatrix}
		\vec{T_1}\\
		\vdots   \\
		\vec{T_N}\\
	\end{pmatrix}\;,
\end{align}

where $\vec{T_n}$ is the filter transmission at each wavelength. Here we used a normalized Gaussian spectral filter with a central wavelength, $\lambda_{n}$, and a full width at half maximum, $FWHM_n$, i.e.\ 
\begin{equation}
	\label{equation: Gaussian filter}
	\vec{T_n} = e^{\frac{\sqrt{2\ln2}(\vec{\lambda}-\lambda_{n})^2}{FWHM_n^2}}
\end{equation}

The reconstructor is described by the following equation: 

\begin{equation}
	\label{equation: reconstruction}
	\mathbf{R}\vec{y} = \hat{\vec{x}}
\end{equation} \noindent

where $\mathbf{R}$ is the reconstruction matrix to obtain the approximation of the original spectrum: $\hat{\vec{x}}$.

\begin{figure}
	\includegraphics[width = \textwidth]{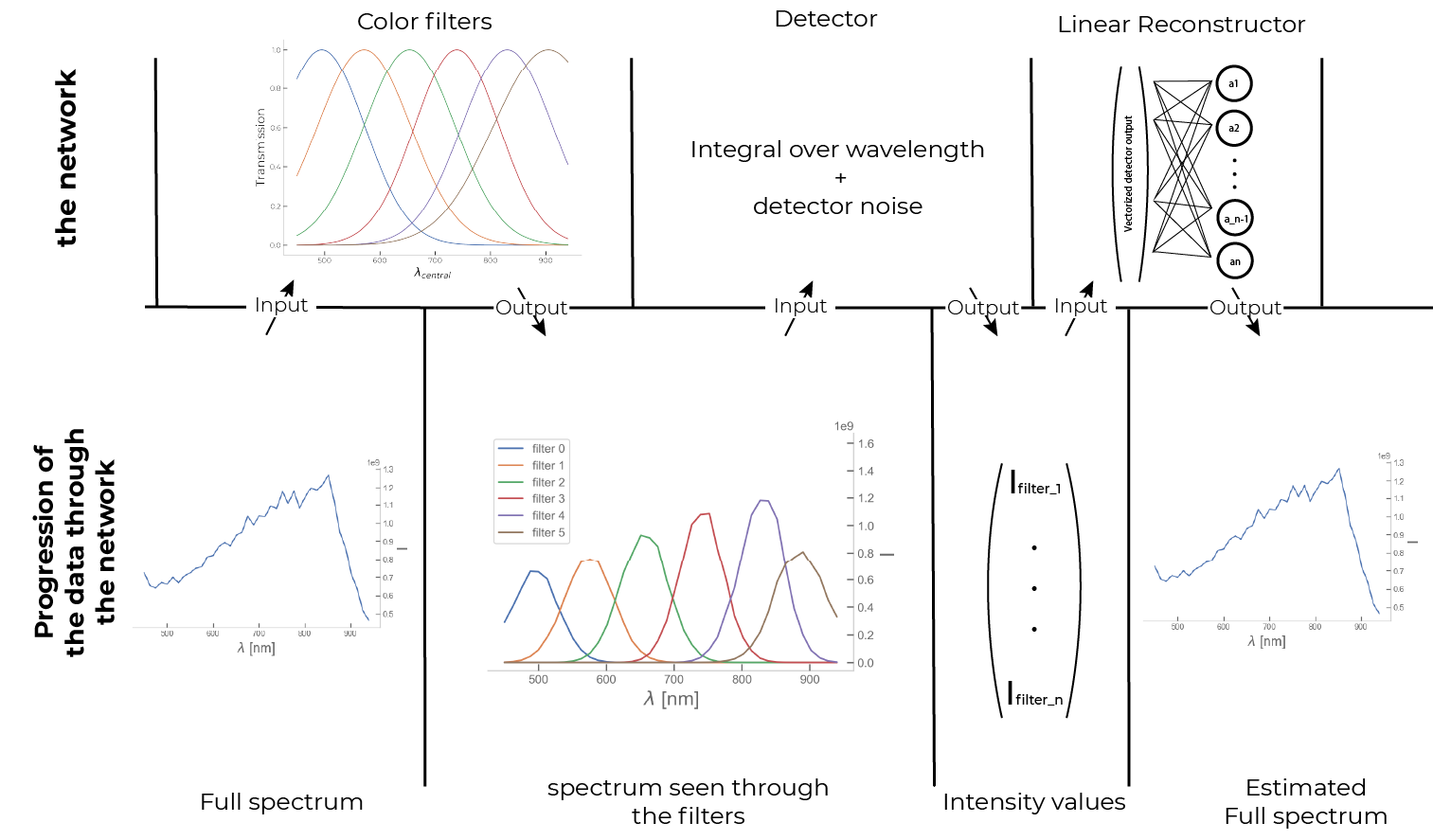}

	\caption{Schematic overview of the optimal filter estimator and the propagation of the data through its linear transformations. }
	\label{figure: bestfilters overview}
\end{figure}

This component is visualized in figure \ref{figure: bestfilters overview}.

\noindent


\subsection{Optimal Layout estimator}
\label{section: optimal layout estimator}
\begin{figure}
	\includegraphics[width = \textwidth]{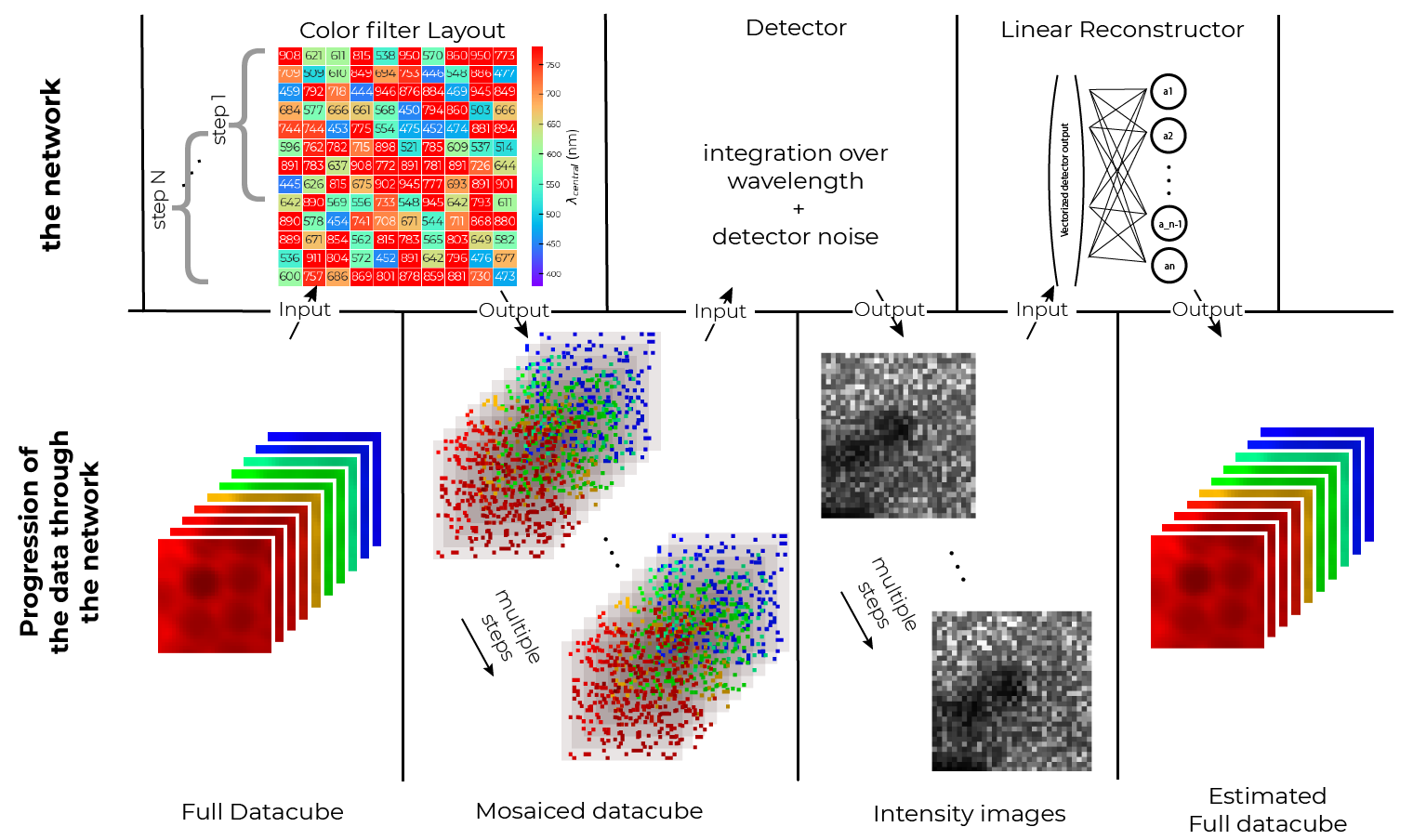}

	\caption{Schematic overview of the network and the propagation of the data through it. It starts with a scene that passes through the spectral filter layout. Some information is being blocked by the filters resulting in a mosaiced cube. This process is repeated for the total number of steps, which are taken in a push broom fashion. The detector flattens the mosaiced hyperspectral cubes into 2D intensity measurements and adds noise. The multiplication of these 2D intensity images with the linear reconstructor results in an estimate of the original data cube.}
	\label{figure: network overview}
\end{figure}

The optimal layout estimator is similar to the optimal filters estimator described above (see Fig.\ref{figure: network overview}. 
It consists of a \textit{layout of spectral filters } and a linear reconstructor that carries out the demosaicing and reconstructs the full hyperspectral data cube. The measurements, where individual pixels see the scene through different spectral filters, can be done either in a snapshot mode or in a push-broom fashion. In the former, only a single-intensity image is acquired. In the latter, the filters are shifted step-wise in one direction across the scene, and multiple images are taken; every step corresponds to one full image carrying different spectral information for all ground pixels.

The optimal layout estimator supports different configurations, which can be grouped into two main configurations: in the first configuration, the spectral filters are fixed and cannot be updated by the algorithm. In the second configuration, however, the filters can be updated as well. This will be further explained below in section \ref{section: Tensorflow implementation}.

Mathematically, we can again describe this component in the compressed-sensing format, referring to equations \ref{equation: CS}-\ref{equation: best filters matrix}. However, $\vec{T_{s,m}}$ now describes the transmission of the filter focused on every geometrical pixel of the scene in every scanning frame, instead of every filter as in equation \ref{equation: Gaussian filter}. Assuming a detector with $M$ pixels and taking $S$ steps, $\mathbf{H}$ can be written as

\begin{align}
	\mathbf{H} &= 
	\begin{pmatrix}
		\vec{T_{11}}\\
		\vdots   \\
		\vec{T_{1M}}\\
		\vec{T_{21}}\\
		\vdots   \\
		\vec{T_{SM}}\\
	\end{pmatrix}\;.
\end{align}

$\vec{x}$ is a serialized version of $S$ times all spectra of all pixels. $\vec{y}$ contains the intensities on the detector of every pixel in every step. The linear reconstruction is the same as above in equation \ref{equation: reconstruction}, but with this different $\vec{y}$. 

\subsection{TensorFlow implementation}
\label{section: Tensorflow implementation}
InSPECtor is implemented in TensorFlow, which provides rapid optimization of all free parameters of our analytical model that is fully differentiable \cite{abadi_tensorflow_2016}. Each component is implemented with a sequence of so-called Layers. A Layer consists of an input of Tensors, an output of Tensors, a differentiable function, and the values of the variables (weights) used in this function along with the input Tensor. Tensors are the main multidimensional data containers of TensorFlow. The physical process represented by each Layer is described by the function and the weights, and the data propagate through each Layer as the inputs and outputs.

Each of these Layers can be made trainable where the values of the weights can be updated by the optimizer, in contrast to remaining fixed during the training phase. The optimizer uses back-propagation to update the weights. Pieces of the training data set are fed to the network, and the resulting output is compared to the desired result using a loss function, in our case the MSE. The loss function is related to every parameter in the framework in the form of a partial derivative. Using that partial derivative, the value of each parameter is updated by the optimization algorithm, to minimize the loss function. In our case, we make use of the adam optimizer \cite{kingma_adam_2017} with different learning rates. The learning rate is a hyperparameter that is found by trial and error. 

The TensorFlow model for the optimal filters estimator consists of three sequential Layers. First is a spectral filter and detector Layer, followed by a noise Layer and, finally, a reconstruction Layer. The second component, the layout estimator, is realized with three TensorFlow Layers. These are a spectral filter layout and detector Layer, again followed by the noise Layer and the reconstruction Layer. 
Each of these Layers is described in more detail below. In addition, we discuss further possible additions to the model, called regularizers.

\subsubsection{Spectral filters and detector}
We describe the filter of each pixel as in equation \ref{equation: Gaussian filter}, a normalized Gaussian profile characterized by a central wavelength and bandwidth. These two numbers make up the weights of this custom spectral filter Layer that can be optimized. The single-spectrum input is multiplied by all the filters. The detector part is simply an integration over wavelength, resulting in an intensity value for each filter. 

The weights corresponding to the central wavelength and bandwidth, respectively, of the spectral filter Layer are scaled to the -1 to +1 range to accelerate training.

\subsubsection{Spectral filter layout and detector}
This Layer is very close to the spectral filter and detector Layer described above. However, the input is a full hyperspectral data cube with the same spatial dimensions as the detector. Each detector pixel has its own filter associated with it, and the multiplication of the filter happens with the spectrum of geometrical pixel imaged on it. Again an integration over wavelength happens to result in one intensity image. For each additional push broom step; this process is repeated with the filters shifted one pixel row with respect to the spatial sampling points, and the detector images are concatenated into a single, long vector.

Different configurations can be implemented with the optimal filter layout estimator. Most of these configurations have an influence on the weights of this Layer. If the configuration contains a fixed layout, the weights of this Layer will be the $\lambda_c$ and $FWHM$ of each filter of the fixed layout. However, if the configuration does not contain a fixed layout, the weights of this layer will be the $\lambda_c$ and $FWHM$ of each filter on each single detector pixel.


\subsubsection{Noise}
We have added a Gaussian noise Layer, which affects the intensity measurements coming from the preceding Layer. The amplitude of this added noise is comparable to the SNR ratio of the input data as noted in \cite{esposito_-orbit_2019}. This layer also ensures the robustness of the design to the physical detector noise and mitigates overfitting. Overfitting denotes fitting not only the underlying patterns in the training data but also the random noise patterns, which will be different and unpredictable when validating the design with different data.

\subsubsection{Reconstruction}
The intensity values from the noisy detector are reconstructed in the final Layer as either a full spectrum or as the full hyperspectral data cube. The reconstruction is implemented as a linear reconstructor. This corresponds to a single Dense Layer in TensorFlow with as many weights as there are entries in the reconstructed spectrum or hyperspectral cube; all bias weights are fixed to 0, to ensure strict proportionality between the input measurement and output data cube. This Dense Layer is initialized with zeros instead of the more common random numbers to help the network converge. To determine the spectrum of one geometrical pixel in the optimal filter layout estimator, the reconstruction can, in theory, make use of all the measurements of all geometrical pixels. However, in practice, it will focus on the connections that contain the most information, e.g. the closest ones. 

\subsubsection{Regularizers}
\label{section: regularizers}
Next to the loss function described above, which compares the output of the network with the desired output, additional loss terms, called regularizers, can be added to the TensorFlow model. Each added loss term must be differentiable as well for it to be able to influence the optimization of the weights. The regularizers can be as important as the network structure itself. Here we apply different regularizers to limit the noise propagation in the linear reconstructor and to implement the discrete filter selection in a differentiable manner. 

The reconstruction layer can be made less prone to overfitting by adding an L2 regularizer. L2 regularization adds the L2-norm of the weights of the linear reconstructor as a loss function and leads to a preference for smaller weights.

\begin{figure}
	\centering
	\includegraphics[width =0.4\textwidth]{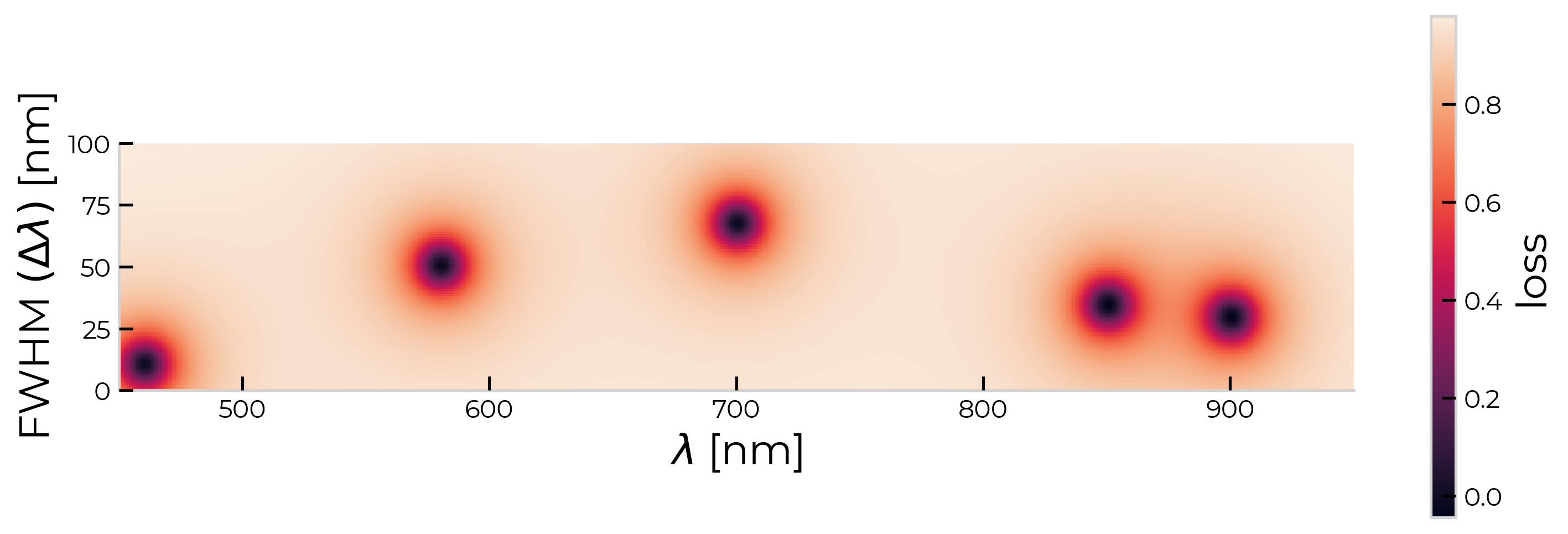}
	\caption{The loss that each filter adds according to its wavelength and FWHM when the following combinations ($\lambda$, $\Delta\lambda$) are desired: (460 nm, 10 nm), (580 nm, 50 nm), (850 nm, 34 nm), (900 nm, 29 nm) and (700 nm, 67 nm). These combinations were randomly selected for display purposes.}
	\label{figure: regularizer_loss_landscape}
\end{figure}

The second regularizer is custom-made to specifically give preference to a fixed selection of filters described by their central wavelengths and widths. This regularizer adds a loss for each filter, but this loss is reduced when the filter resembles one of the selected filters. Since there are two parameters defining each potential passband, the central wavelength and FWHM, the loss function is two-dimensional. For each specified filter, there is a negative 2D normalized Lorentzian function with a global minimum at the specified filter coordinates. A Lorentzian function is preferred over a Gaussian function due to its broader wings, which accelerate convergence. All these Lorentzian functions are summed to create the full loss landscape with local minima at all specified filter coordinates. This sum of Lorentzians is evaluated for each filter in the spectral filter Layer and all these values are added as the additional loss term, which is expressed in the following equation: 

\begin{equation}
	\label{equation: lorentzian2D}
	L = \alpha_{reg} \sum_{pixels} \sum_{filters} 1 - \frac{A^2}{(\lambda - \lambda_{filter})^2+(FWHM-FWHM_{filter})^2+A^2}
\end{equation} \noindent

where $A$ controls how fast a deviation from the desired filter results in a big loss term. $\alpha_{reg}$ determines the weight of this regularizer with respect to the other regularizers and the global loss function. $\lambda_{filter}$ and $FWHM_{filter}$ are the central wavelength and full width at half maximum (FWHM) of the specified filter, respectively. An example of the loss landscape for this regularizer can be seen in figure \ref{figure: regularizer_loss_landscape}. Note that this loss is a unitless quantity with the sole purpose of optimization with respect to certain design constraints; it is not related to any physics in the system. 

The regularizer described above is the key link between the two frameworks of inSPECtor: when the layout is not fixed it ensures that the optimal filters from the optimal filters framework are highly preferred over all others in the layout design. 

\subsubsection{Configurations of the framework}
As mentioned in section \ref{section: optimal layout estimator}, the filter layout estimator can be configured in many ways. The weights in the color-filter Layer and the reconstruction Layer can be set to be trainable or not. In addition, the initialization of the weights of these layers and the choice of regularizers can be selected. 

A configuration of the framework is determined by the following items: 
\begin{itemize}
    \item Determine the initialization of the filters.
    \item Either optimize the filters or fixate the filters to the initialized values. 
    \item Determine the initialization of the layout. 
    \item Either optimize the layout or fixate the layout to the initialized pattern.
    \item If both the filters and the layout need to be optimized, do they need to be optimized simultaneously?     
\end{itemize}

Throughout this paper, we will initialize the filters in the same way. We will call this initialization "regular filters". The term "regular filters"  represents filters with identical Gaussian passbands, spaced in wavelength by their FWHM and spanning the complete wavelength range of interest (450-940nm). Two regular filters would both have a FWHM of 245 nm and be centered at 577.5 nm and 818.5 nm, respectively.

When the filters are optimized, there are two possibilities called "best filters" and "optimized filters". They are related to the last item that determines the configuration: the term "best filters" means that the filters are the result of the optimization done by the "Optimal filters estimator", see section \ref{section: optimal filters}. The final term, optimized filters, means that the filters are optimized together with the layout in the "optimal layout estimator", described above in section \ref{section: optimal layout estimator}.

The layout is initialized either randomly or with a fixed pattern. In the section below, the two fixed patterns that we use are described: an LVF-like pattern and the "squarish" pattern. In future work, more patterns could be implemented in this framework as long as they are generated with a specific function that is differentiable and not limited to certain sizes of the simulated detector. 


\subsubsection{Fixed patterns}
We describe two different fixed patterns in this section, an LVF-like pattern and our own "squarish"-pattern.

The LVF-like pattern is a repreating arrangement of the different filters with central wavelengths increasing in the scanning direction, whilst being uniform in the other direction.  

The “Squarish” pattern can be best defined as a lattice in statistical physics, by a unit cell and two linearly independent primitive translation vectors. We keep to the definition 12.2.1 in \cite{simon_oxford_2013} for a unit cell as follows: A unit cell is the repeated motif which is the elementary building block of the periodic structure. 

In our case, a unit cell contains each spectral filter at least once in a fixed layout. For instance, if a 500-nm filter is directly to the right of a 650-nm filter in the unit cell, all 650-nm filters in the total layout will find a 500-nm filter to their right except for the edges of the sensor. 

The unit cell for our “squarish” pattern is defined to have a unit cell that is as close as possible to a square within the square grid graph \cite{acharya_index_1981}. When a perfect square is not possible, the direction perpendicular to the scanning direction is filled first. 

The primitive translation vector is the vector between the same filter in two different unit cells. One of the primitive translation vectors can always be defined to be perfectly aligned perpendicular to the scanning direction. The second is then usually fixed due to the requirement that the pattern must be uninterrupted. There is one exception when the unit cell is a perfect rectangle. In that case, the translation vector is chosen to be never perfectly aligned with the scanning direction, but always skewed by one pixel. This skew is introduced to ensure that a full cycle of all filters is repeated in the scanning direction, instead of a repetition of a subset, as in a Bayer pattern. One example is shown in figures \ref{figure: squarish fixed detector} and \ref{figure: squarish joint detector}

\subsection{Training, validation and test data}
\subsubsection{Hyperscout data}
\begin{figure}
	\centering
	\includegraphics[width=0.6\textwidth]{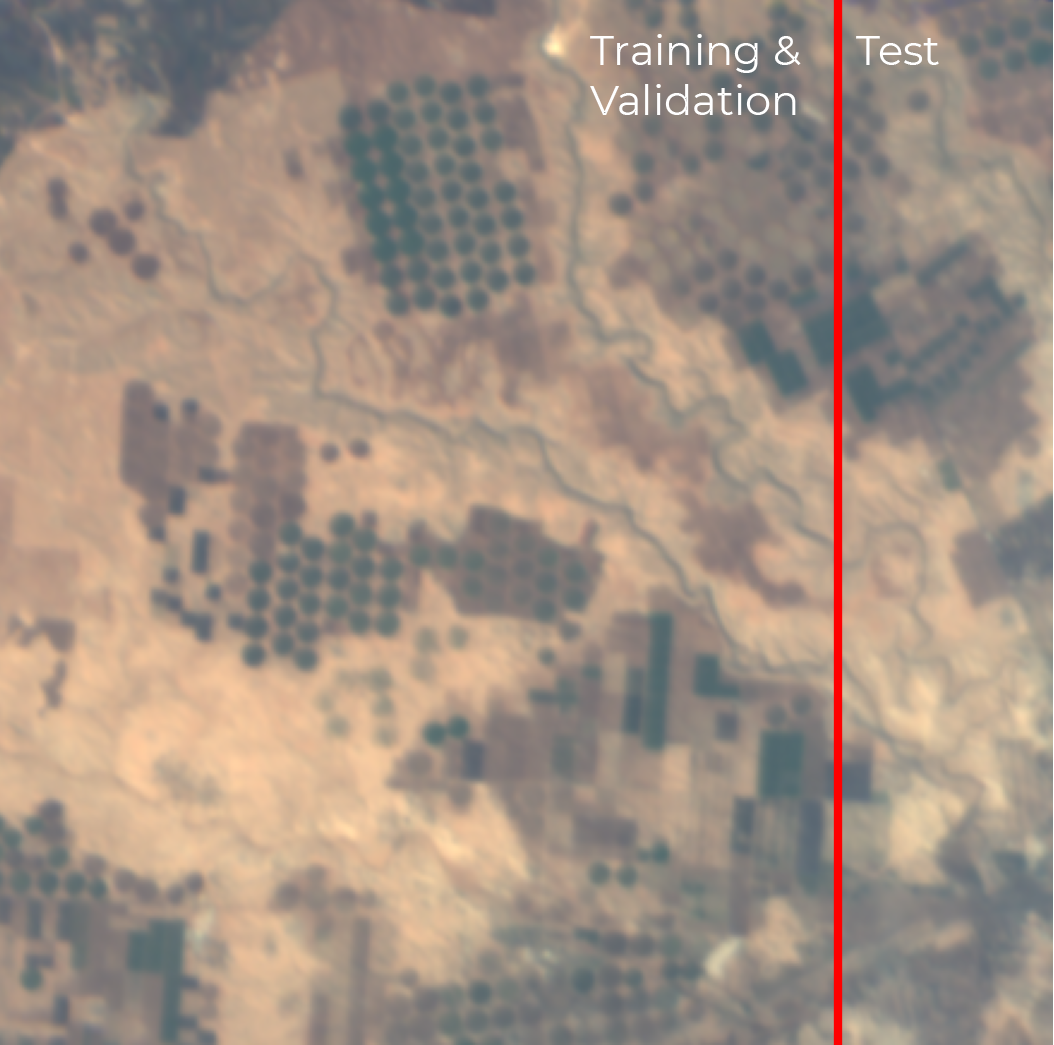}
	\caption{An RGB representation of the sampled scene. The red line shows the divide between the test set on the right and the training and validation set on the left.}
	\label{figure: Hyperscout RGB}
\end{figure}

To train and validate the framework, we have made use of satellite data obtained by the Hyperscout instrument \cite{esposito_-orbit_2019-1}. We used one of the 440 by 440-pixel images with a ground sampling distance of 70 m. There are 40 wavelength bands spanning 450-940 nm with a spectral resolution of about 15 nm each. The scene is shown in Fig.\ref{figure: Hyperscout RGB} as an RGB picture and contains agricultural fields as well as sand, rivers, and clouds. The RGB picture was generated by using the Python package "colour science" and scaling the colors to google maps satellite imagery. 

The satellite data was separated in 2 different parts as shown in Fig.\ref{figure: Hyperscout RGB}. 350 columns, 154.000 spectra, were reserved for training (training set) and the hyperparameter selection (validation set). The separate test set consisted of the remaining 90 columns, or 39.600 spectra. This test set was kept separated during training and the selection of the best performing hyperparameters and was only used to present the final results of the network noted in the sections below.  

The optimal filters were determined by training and validating on 100,000 randomly selected spectra from the training \& validation part of the Hyperscout data set. The resulting filters are subsequently tested on 10,000 randomly selected spectra from the test set. 

For the optimal layout estimator, we transformed the Hyperscout data set into patches of 10x10 pixels. This size corresponds to the used detector size of 10x10 physical pixels. The training data and validation data consisted of 10,000 patches from the corresponding part of the Hyperscout data set that were randomly augmented by mirroring and/or rotation by 90, 180, or 270 degrees. Although there is a high likelihood that there is a partial overlap between some different patches, exact duplicates were avoided. The test set consisted of 1000 patches without any augmentation from the test part of the Hyperscout data set. 

To separate between the training set and validation set, we used the inbuilt separation function of TensorFlow. This randomly selected 10\% for the calculation of the validation loss and 90\% for the actual training of the network.

\subsubsection{Information content of the data}
\label{chapter: data exploration}
The power of compressed sensing lies in using the correlations in space and wavelength of the scene. In this section we determine the information content of the used data to estimate the amount of compression that will be viable. To this end, we 1) assess the spatial correlations at a given wavelength with a Fourier analysis and the spatial and spectral correlations by calculating the PSNR between pixels as a function of their distance and 2) determine the information contained in the spectra with a Principle Component Analysis (PCA).

\begin{figure}
	\centering
	\begin{subfigure}[t]{0.47\textwidth}
	\includegraphics[width = \textwidth]{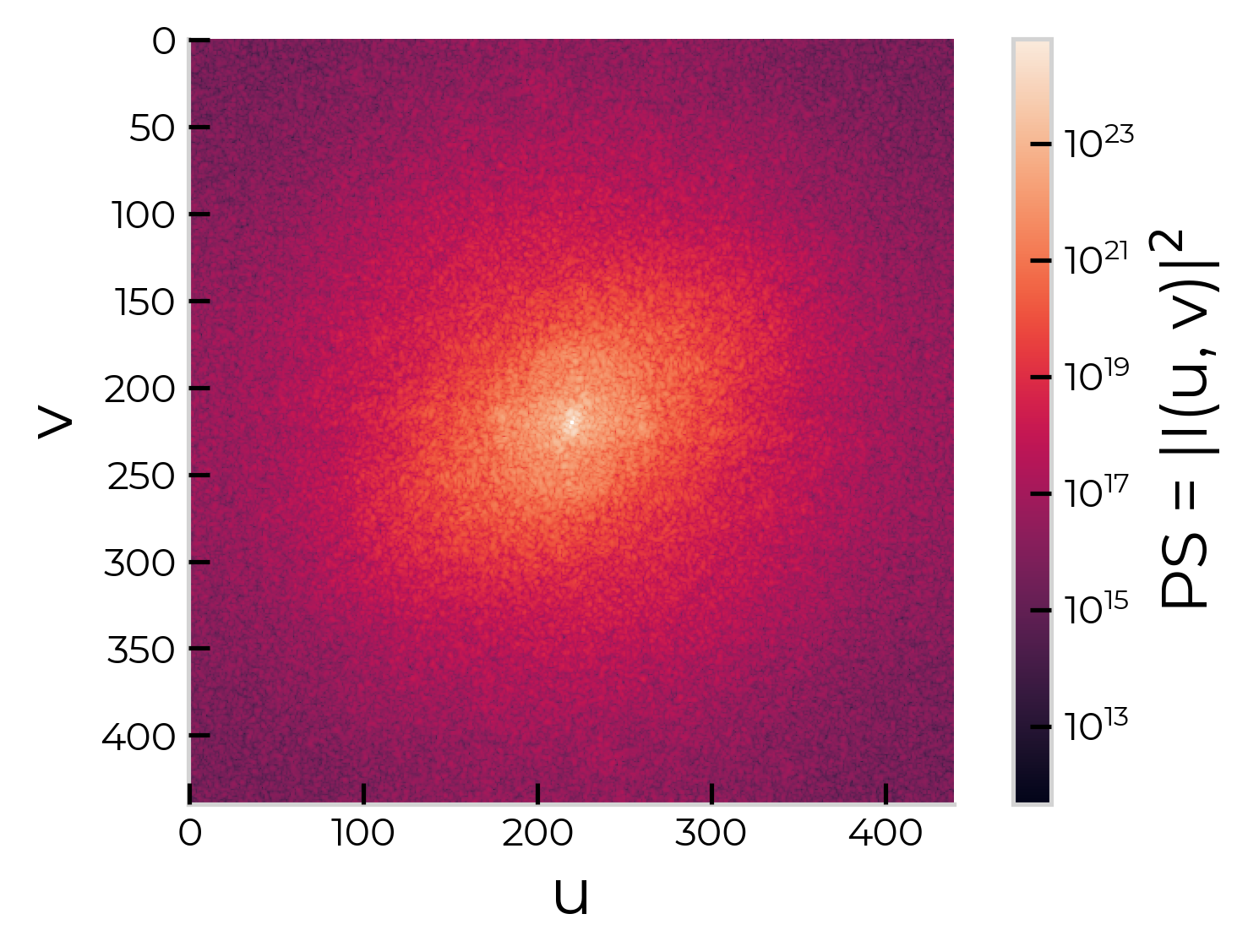}
	\caption{The power spectrum of the image at a wavelength of 701 nm, tapered with a Bartlett-Hann window towards the average.}
	\label{figure: PS}
	\end{subfigure}
	\hfill
	\begin{subfigure}[t]{0.47\textwidth}
	\includegraphics[width = \textwidth]{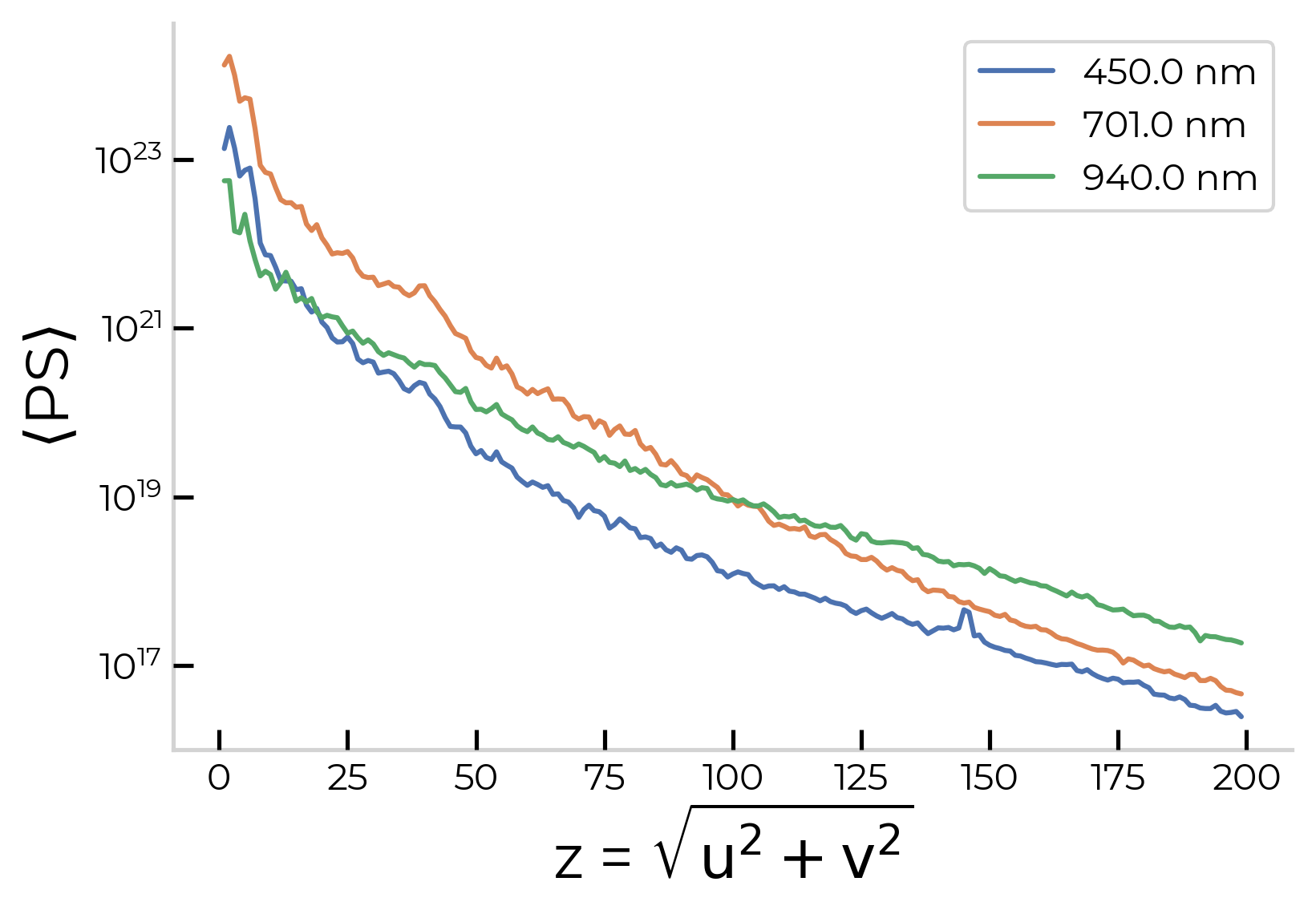}
	\caption{The azimuthal average of the power spectrum at different wavelengths.}
	\label{figure: PS_z}
	\end{subfigure}
	\caption{Fourier analysis}
	\label{figure: Fourier analysis}
\end{figure}

The simplest method to analyze the spatial correlations is the power spectrum of the image at a given wavelength (see figure \ref{figure: Fourier analysis}). Figure \ref{figure: PS} shows that the power spectrum is approximately azimuthally symmetric, which allows us to limit ourselves to the azimuthal average in different wavelengths (see figure \ref{figure: PS_z}). We observe a gradual decline with increasing spatial frequency; the data set does not show any flattening at the higher frequencies, which indicates that the data are not dominated by white noise even at the highest spatial frequencies.

A flattening of the power spectrum would indicate that the pixel binning is too high for the resolution of the optical system, and neighboring pixels would sample the same resolution element. In the absence of this flattening, we conclude that the system is not spatially oversampled.

\begin{figure}
	\centering
	\includegraphics[width = \textwidth]{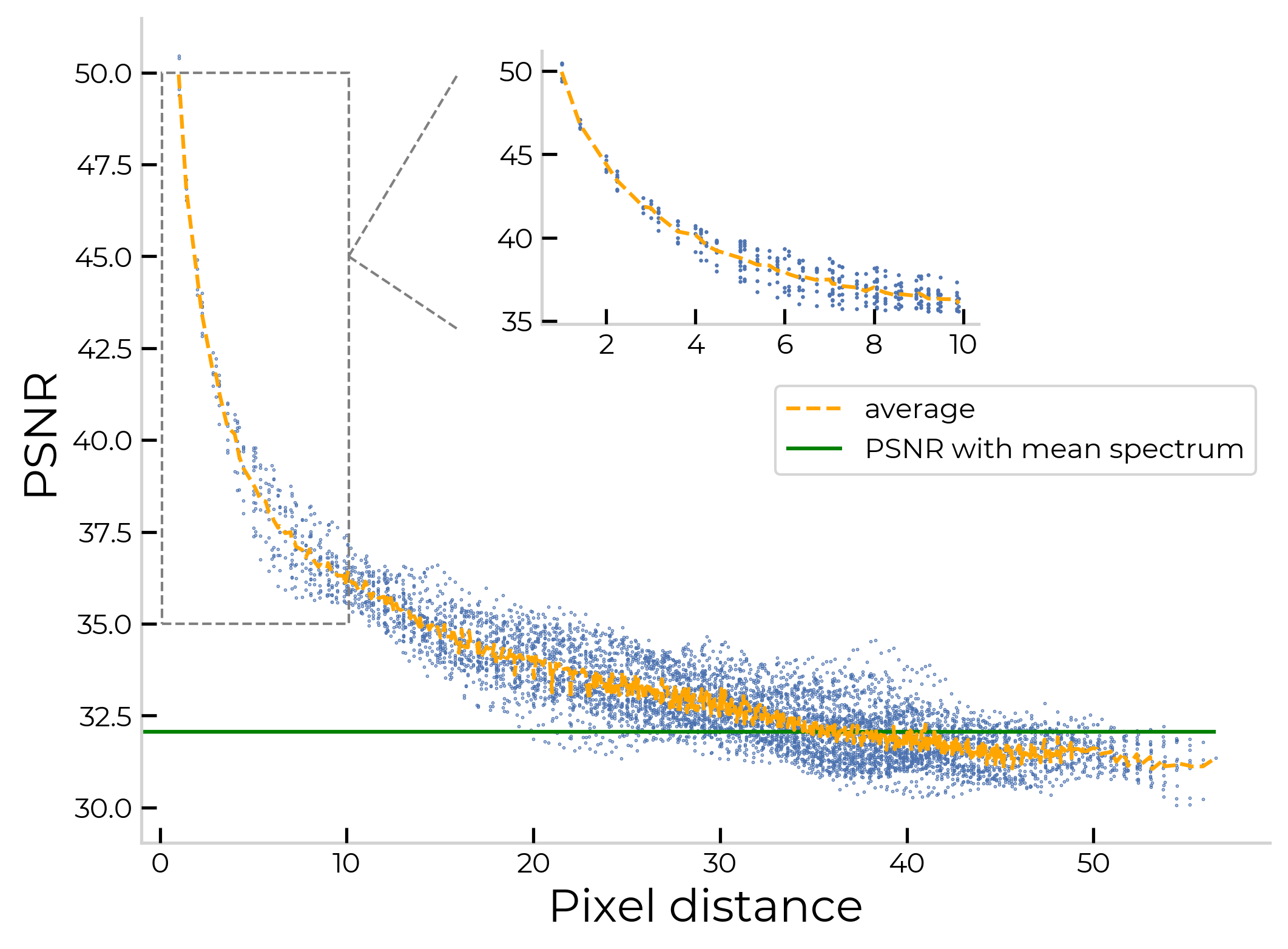}
	\caption{The blue points show the PSNR between the spectra of two pixels as a function of the distance between them. The orange line is the average of all pairs with the same distance. The green line is the PSNR of a comparison of a pixel's spectrum to the mean spectrum of the whole image, averaged over all pixels. A close-up of the PSNR for distances of 10 pixel and less is shown in the upper right.}
	\label{figure: PSNR_distance}
\end{figure}

To analyze the spectral correlations in space, we determined the average PSNR of two pixels as a function of their distance, which is shown in figure \ref{figure: PSNR_distance}. When pixels are close-by, their spectra are very similar (high PSNR). However, when two pixels are far from each other, their spectra differ greatly. When the distance exceeds 34 pixels, the spectrum of a pixel is better approximated by the mean spectrum of the whole image than the spectrum of a random, far-far-away pixel. This indicates the distance at which all but the most basic correlation between pixels vanishes. 

\begin{figure}
	\centering
	\begin{subfigure}[t]{0.32\textwidth}
	\includegraphics[width = \textwidth]{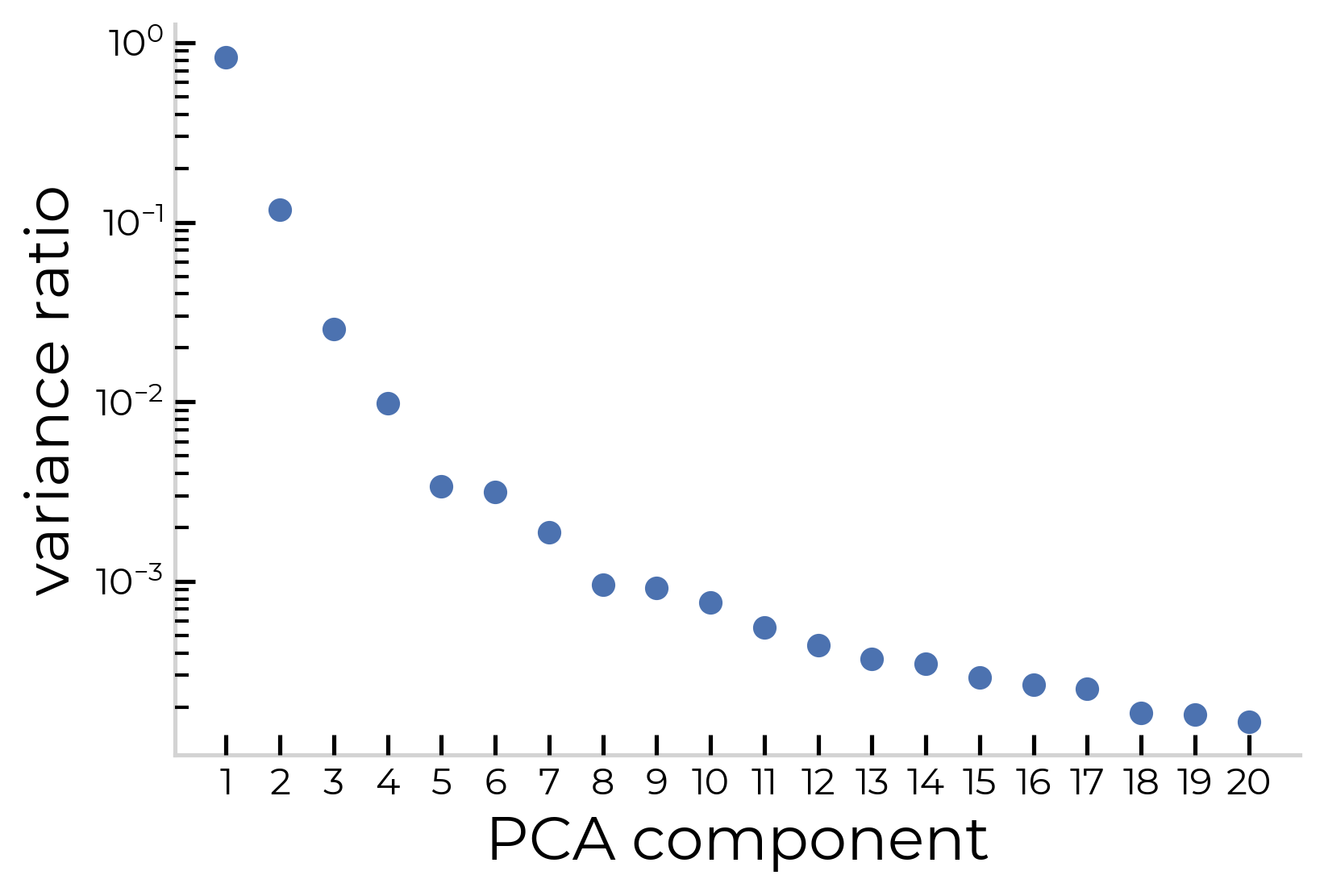}
	\caption{Variances associated with each PCA component of the spectra in the image. The variances are normalized to add up to 1.}
	\label{figure: PCAvariances}
	\end{subfigure}
	\hfill
	\begin{subfigure}[t]{0.32\textwidth}
	\includegraphics[width = \textwidth]{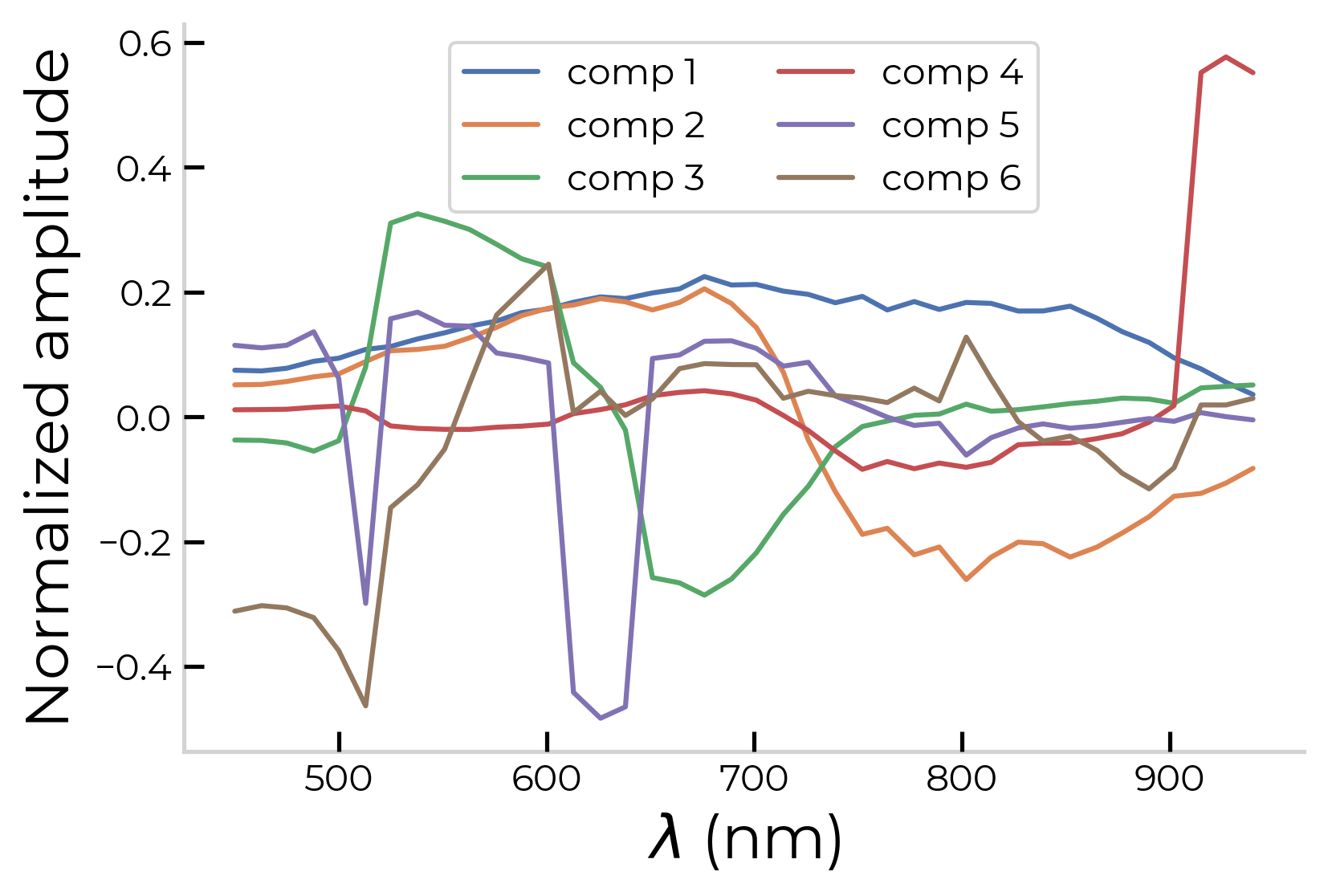}
	\caption{PCA components of the spectra in the Hyperscout data set.}
	\label{figure: PCAcomponents}
	\end{subfigure}
	\hfill
	\begin{subfigure}[t]{0.32\textwidth}
	\includegraphics[width = \textwidth]{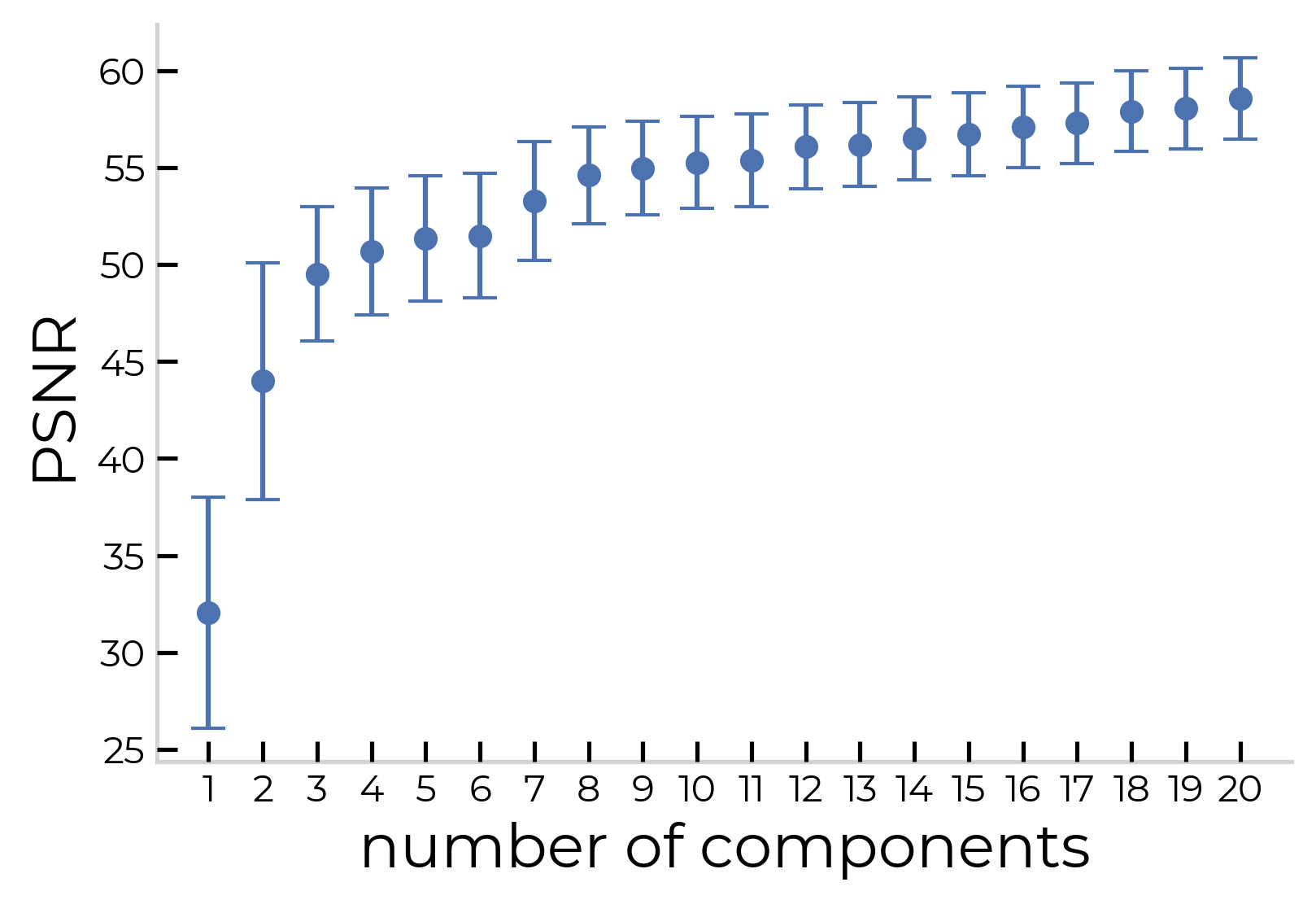}
	\caption{The average PSNR between a spectrum and its approximation as a function of the number of PCA components used for this approximation.}
	\label{figure: PCA_PSNR}
	\end{subfigure}
	\caption{PCA analysis}
	\label{figure: PCAanalysis}
\end{figure}

To assess the information content of the spectra, we carried out a Principal Component Analysis (PCA, see figure \ref{figure: PCAanalysis}). The drop-off in the variance of successive PCA components is very sharp, indicating that much of the spectra can be approximated with a small number of PCA components (see figure \ref{figure: PCAvariances}). The individual PCA components shown in figure \ref{figure: PCAcomponents} seem to pick up the Vegetation Red Edge (component 2) and water vapor absorption in the NIR (component 4).

Finally, we have calculated the average PSNR between a spectrum and its approximation per number of PCA components used for this approximation. This calculation shows much each PCA component adds to reproducing the original data. The number of components that results in an acceptable approximation is related to the number of filters needed for an acceptable reconstruction. However, since there are no negative filters, and interference filter transmission profiles have limitations, the number of PCA components cannot be converted directly to the number of required optical filters.

From four PCA components onwards, the improvements in approximation are minor. Our expectation is that the first PCA component could be approximated by a filter directly. However, for the second PCA component two filters would be necessary: one until 700 nm and one starting at 700 nm. For the third PCA component, we expect a filter from 500 to 600 nm and one from 600 to 700 nm. Finally, the fourth PCA component could be done with a single filter at 900 nm. Adding all these filters, we expect that 7 filters would be necessary to adequately approximate the full spectrum at 40 wavelength bands. 

\section{Results}
In this section, we show the results produced by InSPECtor as a function of number of filters and number of steps. In the TensorFlow environment these two variables are implemented as hyperparameters:
\begin{itemize}
    \item the number of push broom steps to take, going from a snapshot image (one step) to two or more frames 
    \item the number of filters present in the layout, going from two different filters up to 19. 
\end{itemize} 
The PSNR and MSE are evaluated at all permutations of these two hyperparameters.

\begin{figure}
        \centering
	\begin{subfigure}[t]{0.33\textwidth}
	\includegraphics[width = \textwidth]{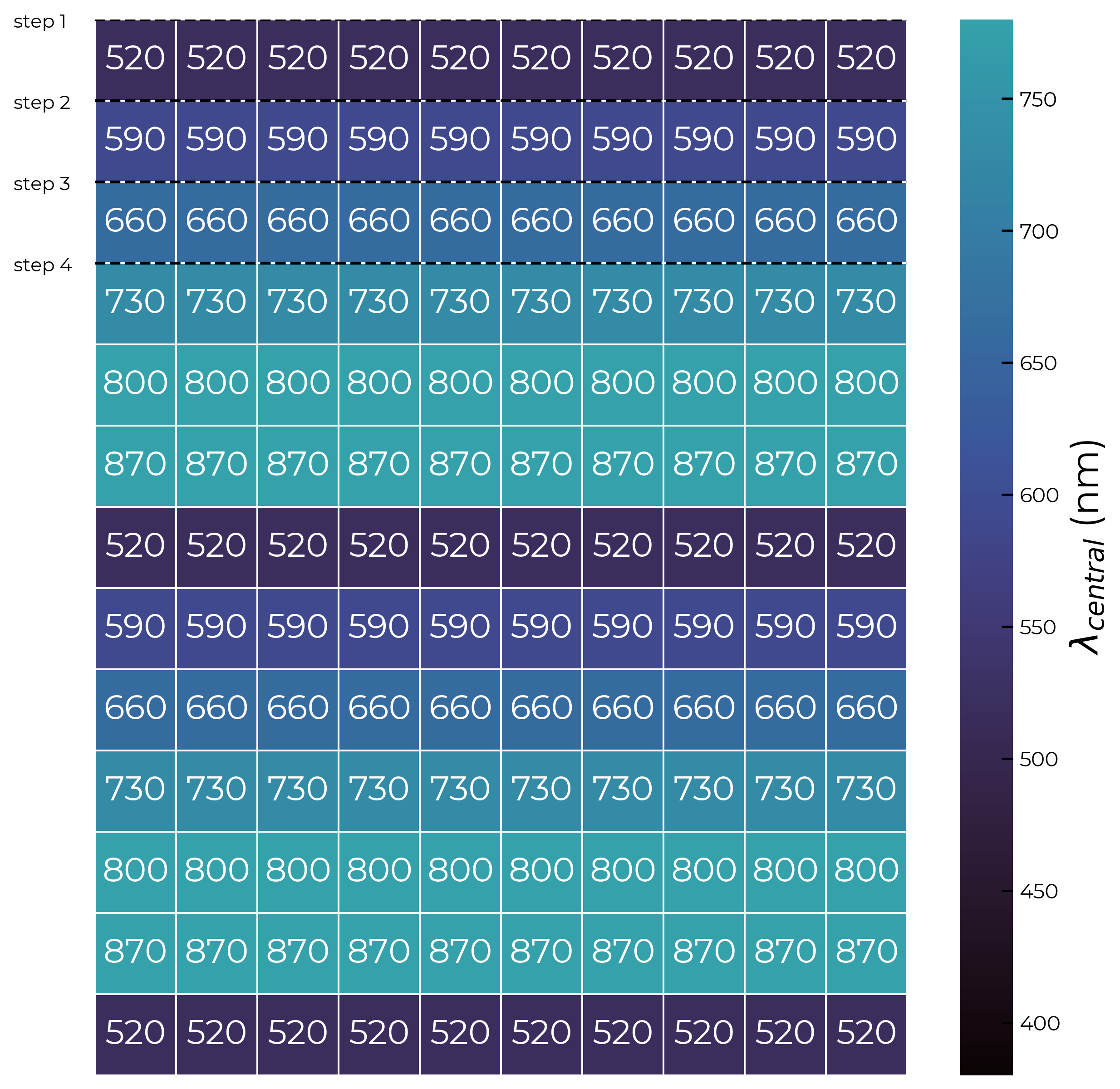}
	\end{subfigure}
	\qquad
	\begin{subfigure}[t]{0.33\textwidth}
	\includegraphics[width = \textwidth]{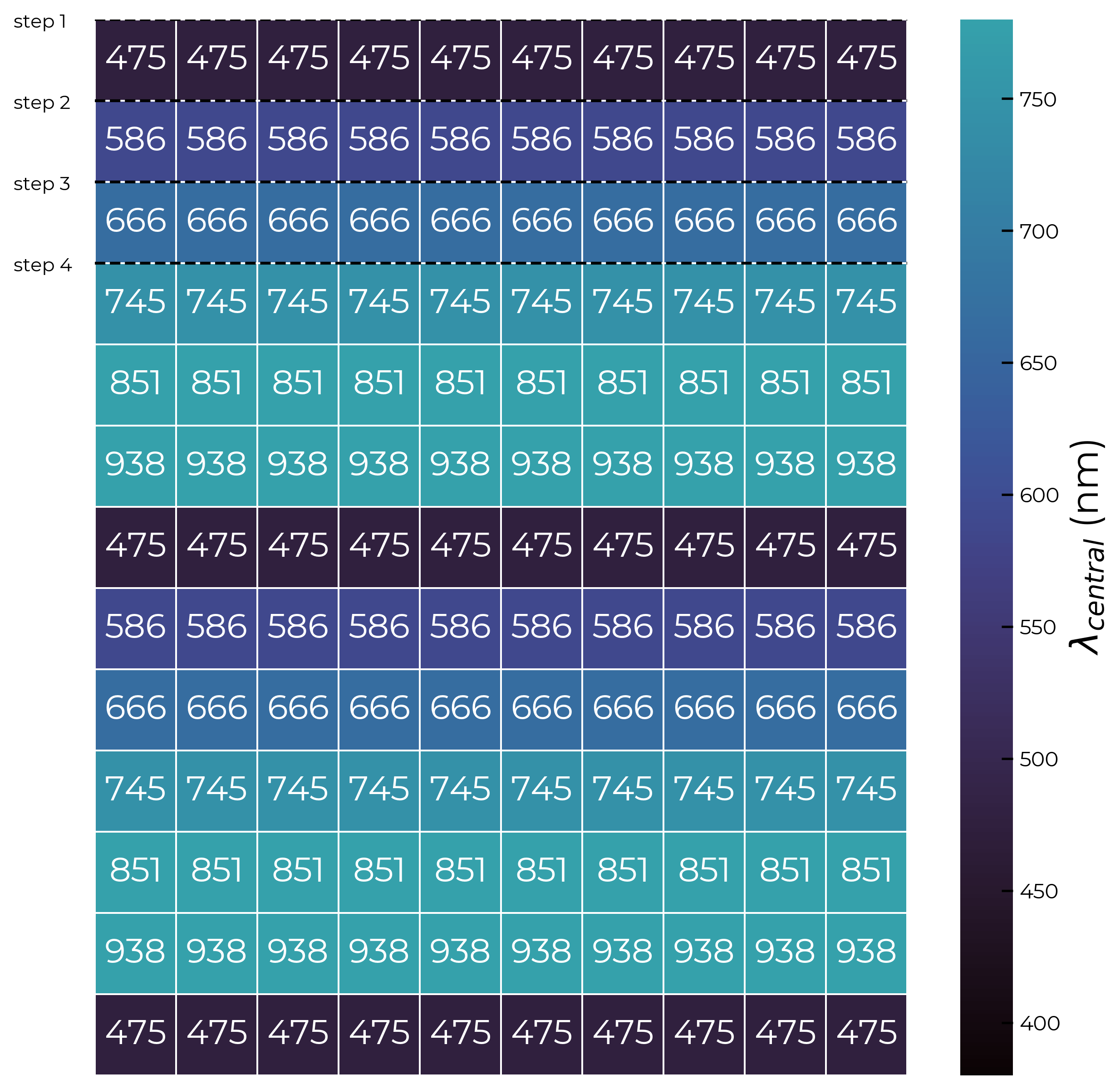}
	\end{subfigure}
 
        \medskip
        
	\centering
	\begin{subfigure}[t]{0.33\textwidth}
	\includegraphics[width = \textwidth]{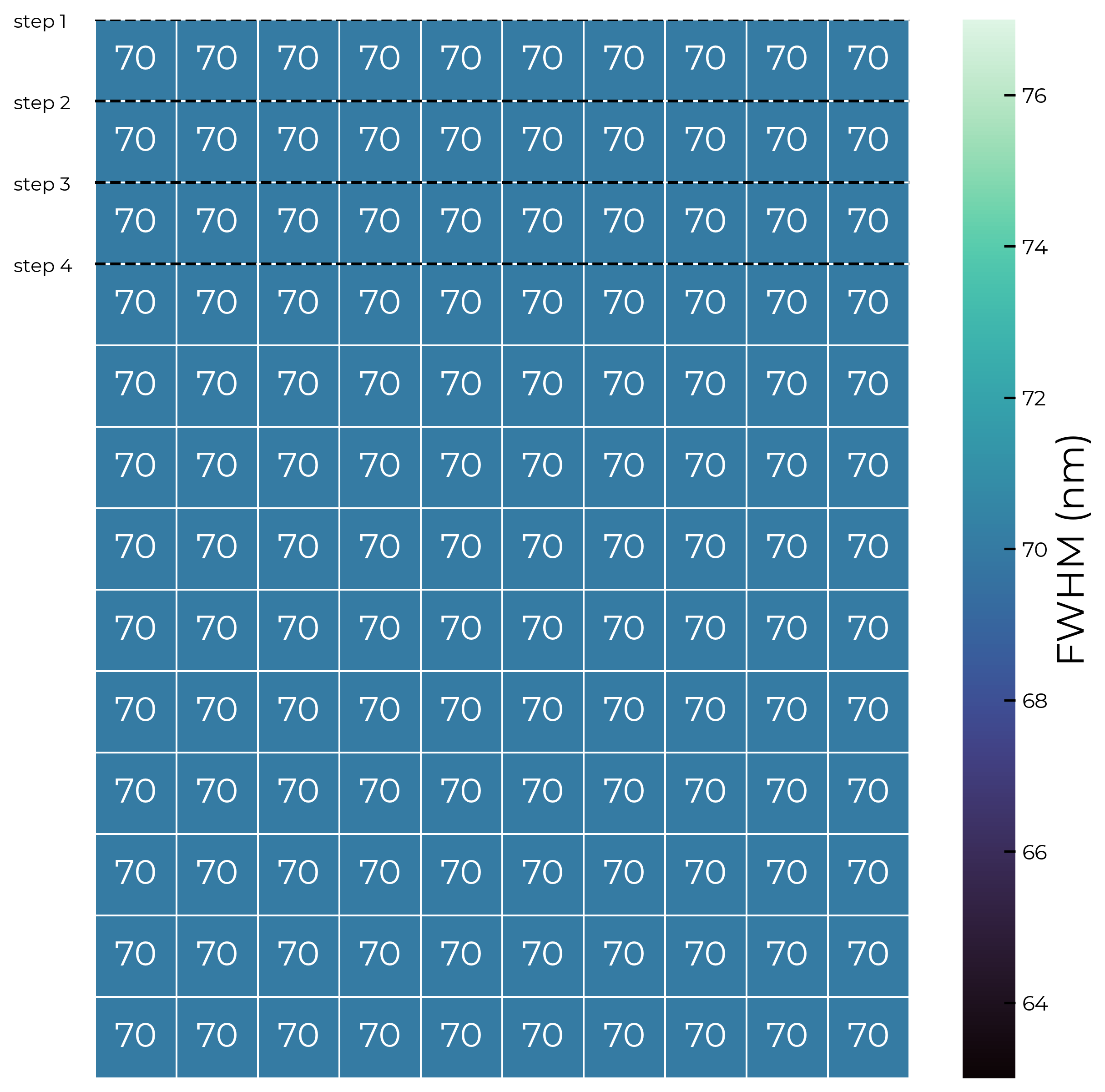}
	\caption{LVF layout, with fixed regular filters.}
	\label{figure: classical detector}
	\end{subfigure}
	\qquad
	\begin{subfigure}[t]{0.33\textwidth}
	\includegraphics[width = \textwidth]{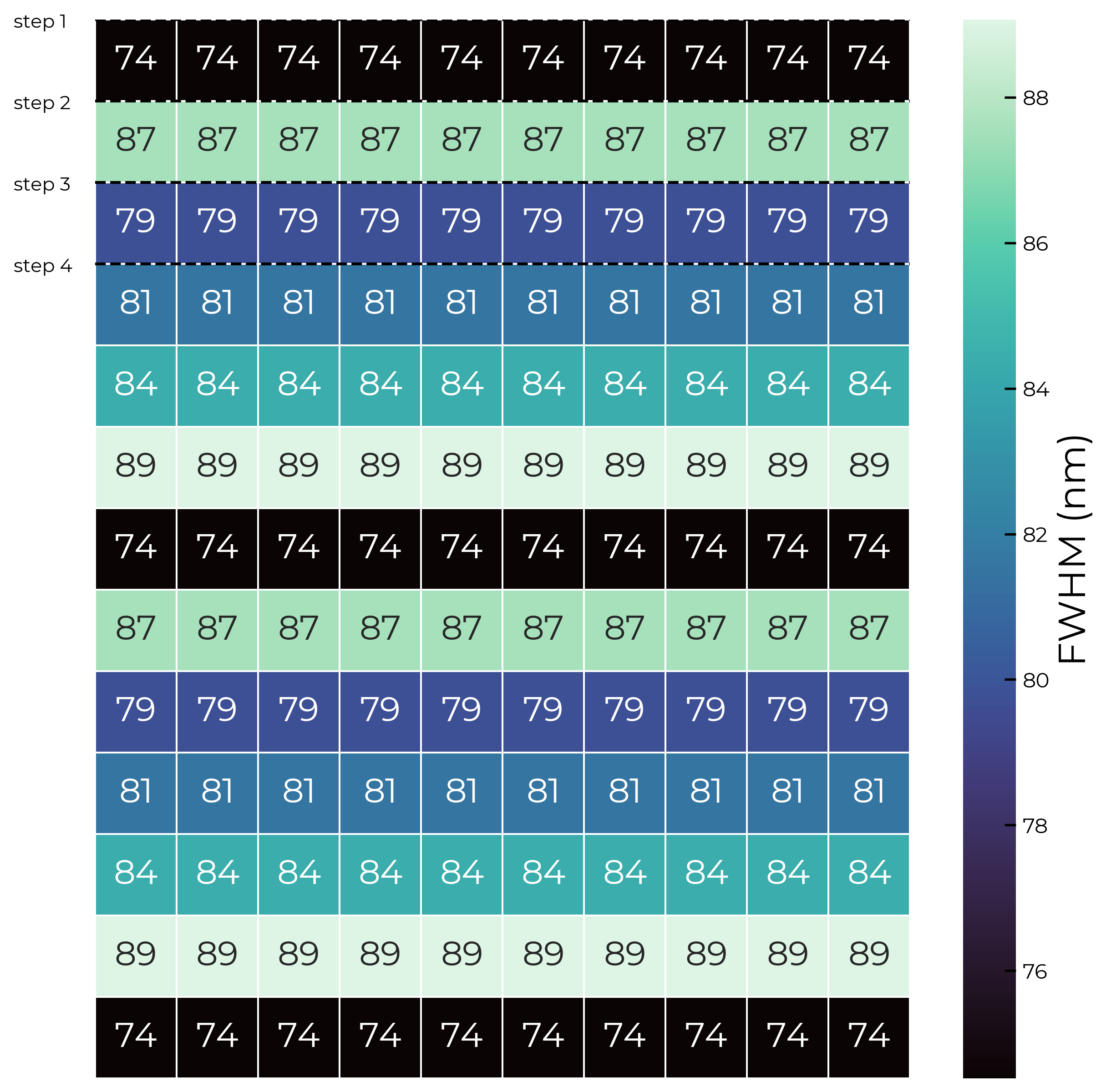}
	\caption{LVF layout with fixed filters separately optimized by the first component.}
	\label{figure: bfregu detector}
	\end{subfigure}

        \bigskip

	\centering
	\begin{subfigure}[t]{0.33\textwidth}
	\includegraphics[width = \textwidth]{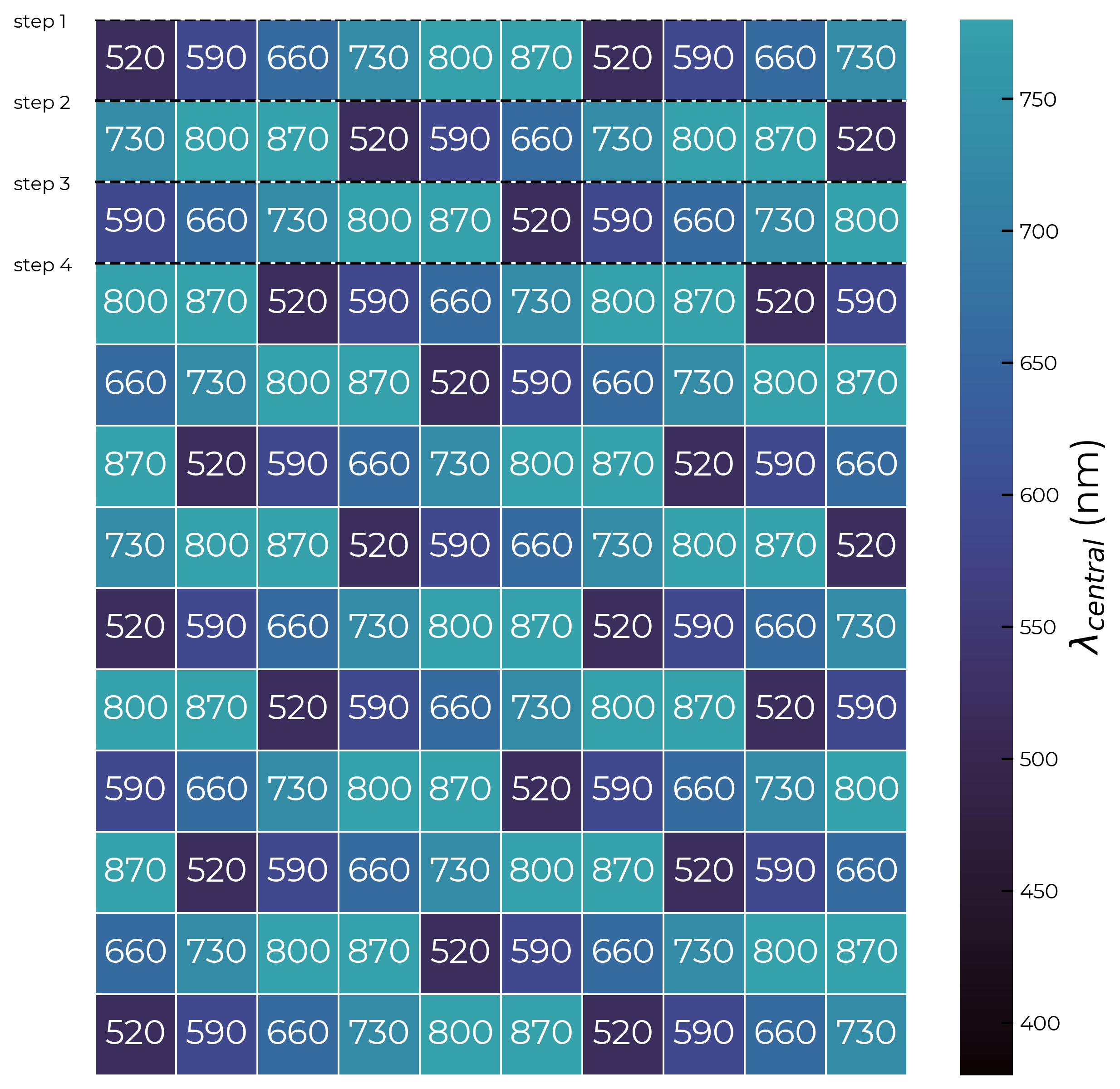}
	\end{subfigure}
	\hfill
	\begin{subfigure}[t]{0.32\textwidth}
	\includegraphics[width = \textwidth]{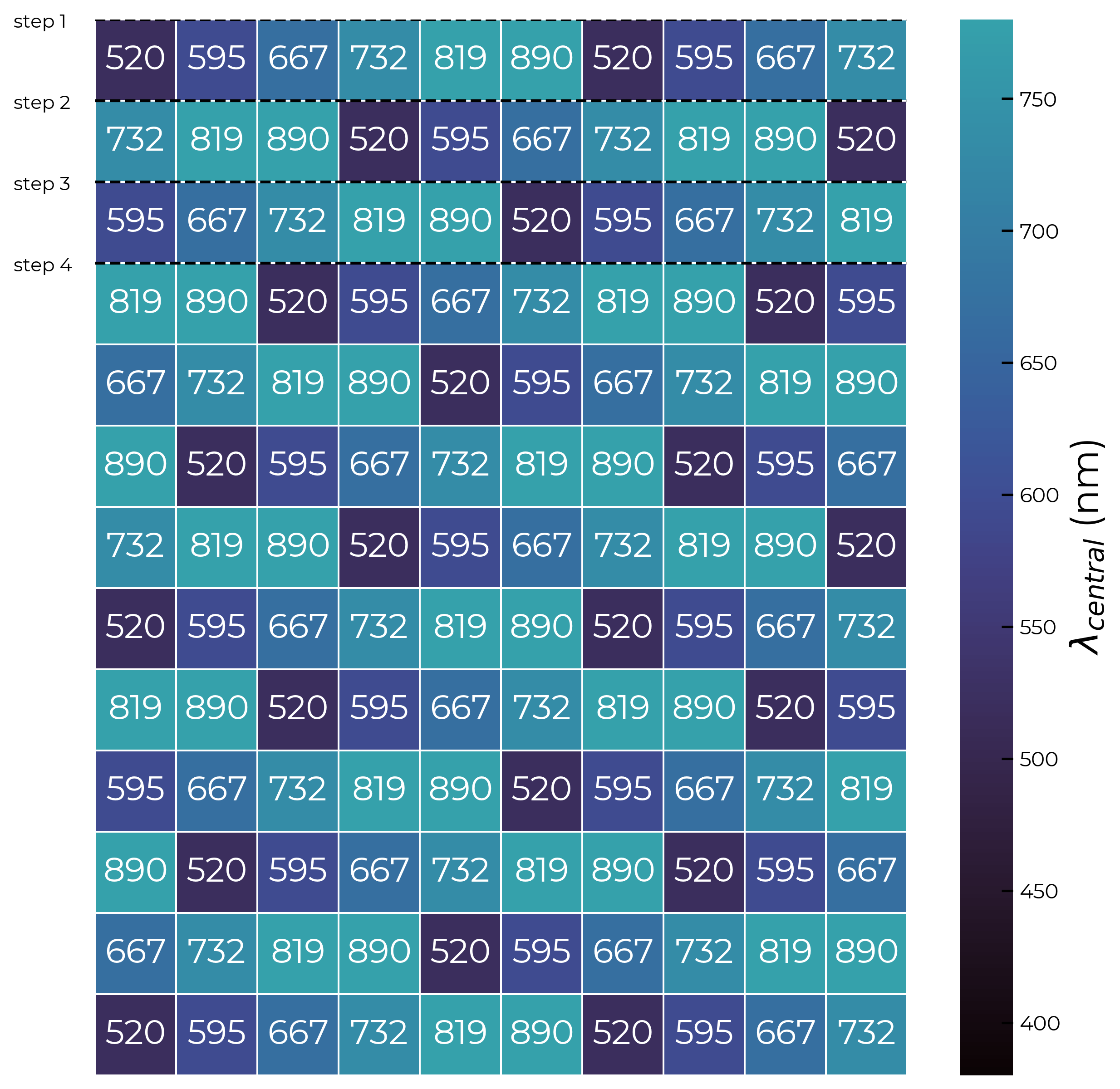}
	\end{subfigure}
	\hfill
	\begin{subfigure}[t]{0.33\textwidth}
	\includegraphics[width = \textwidth]{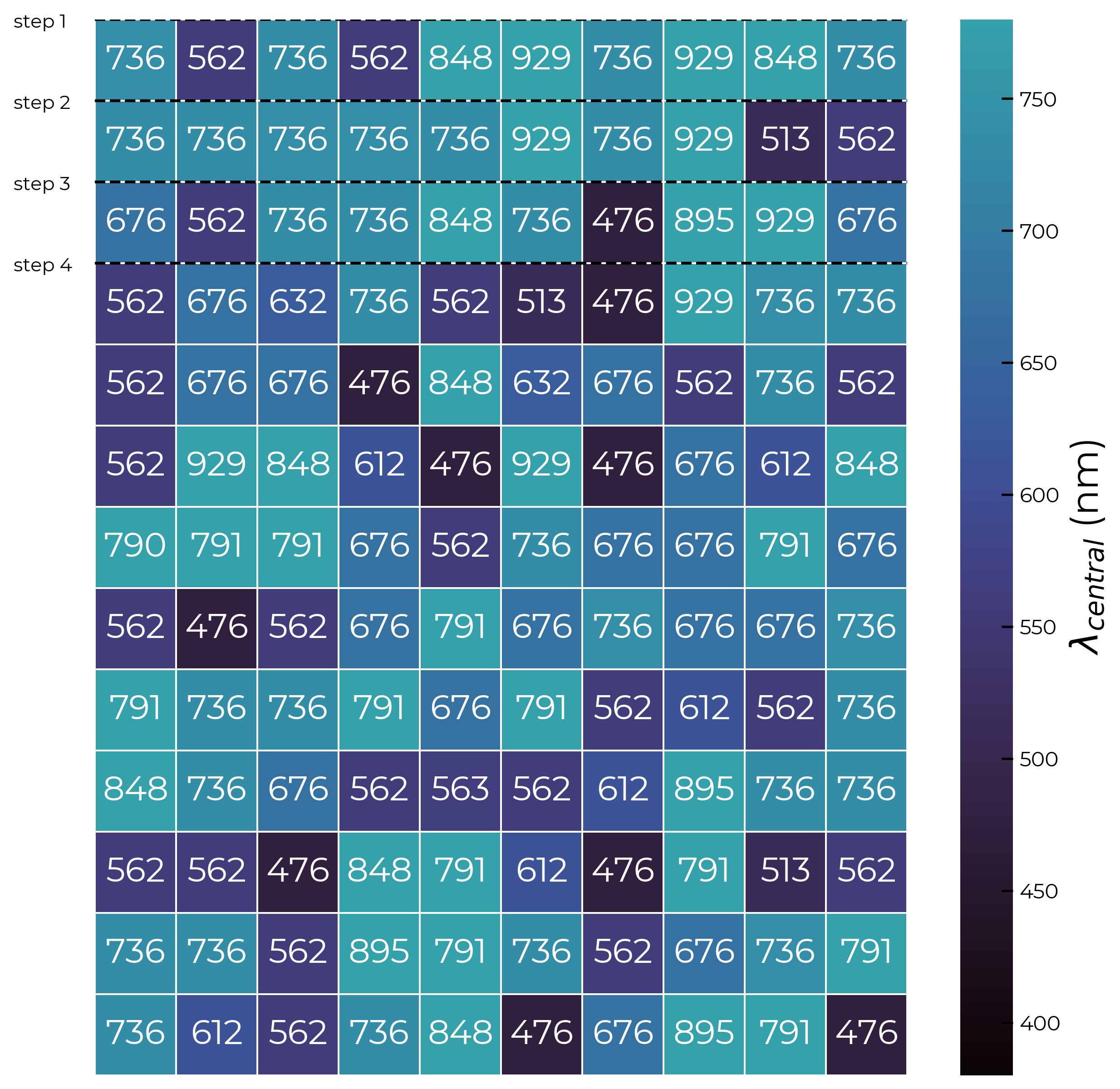}
	\end{subfigure}

	\medskip
	\centering
	\begin{subfigure}[t]{0.33\textwidth}
	\includegraphics[width = \textwidth]{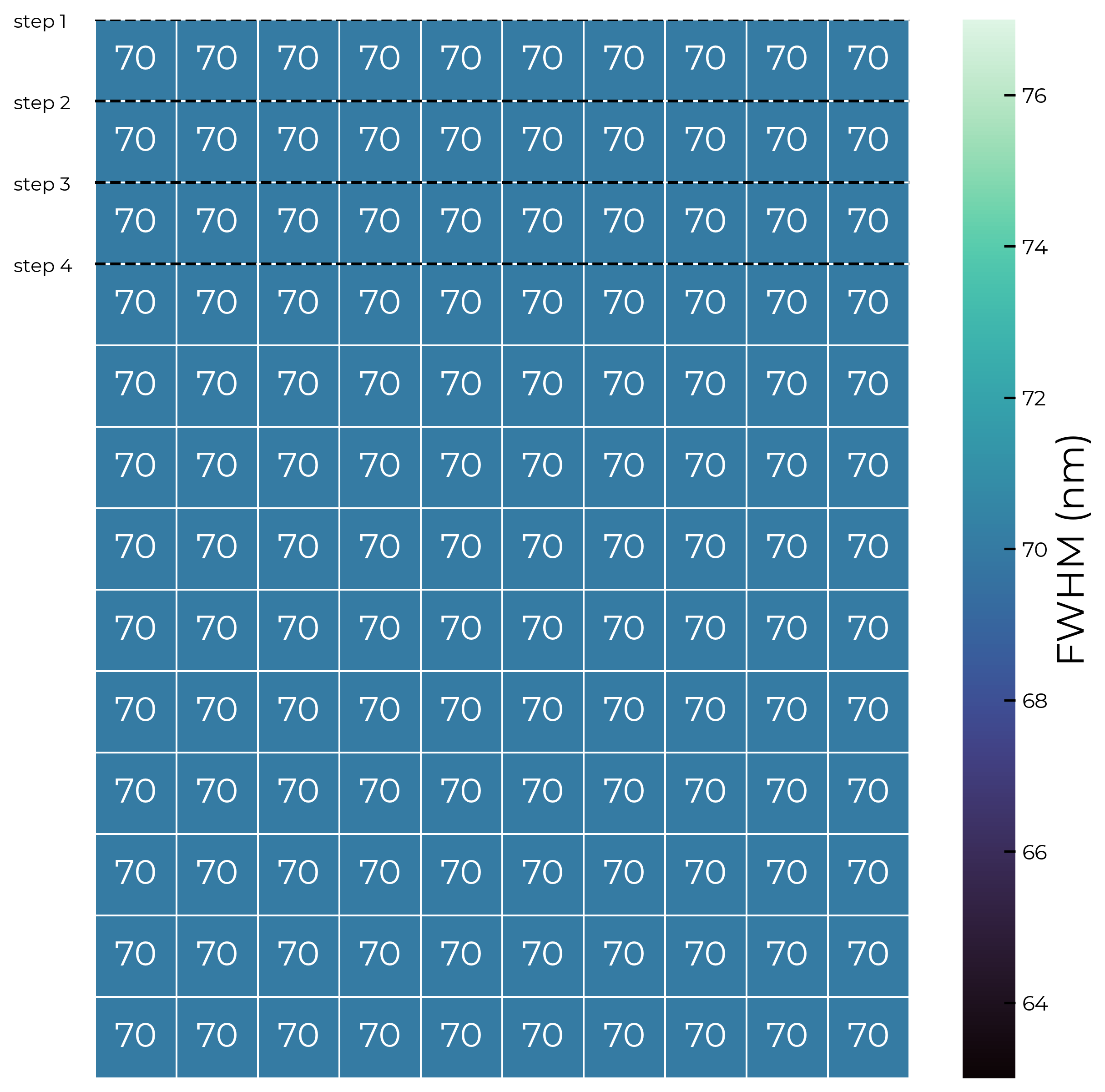}
	\caption{Squarish layout with fixed regular filters.}
	\label{figure: squarish fixed detector}
	\end{subfigure}
	\hfill
	\begin{subfigure}[t]{0.32\textwidth}
	\includegraphics[width = \textwidth]{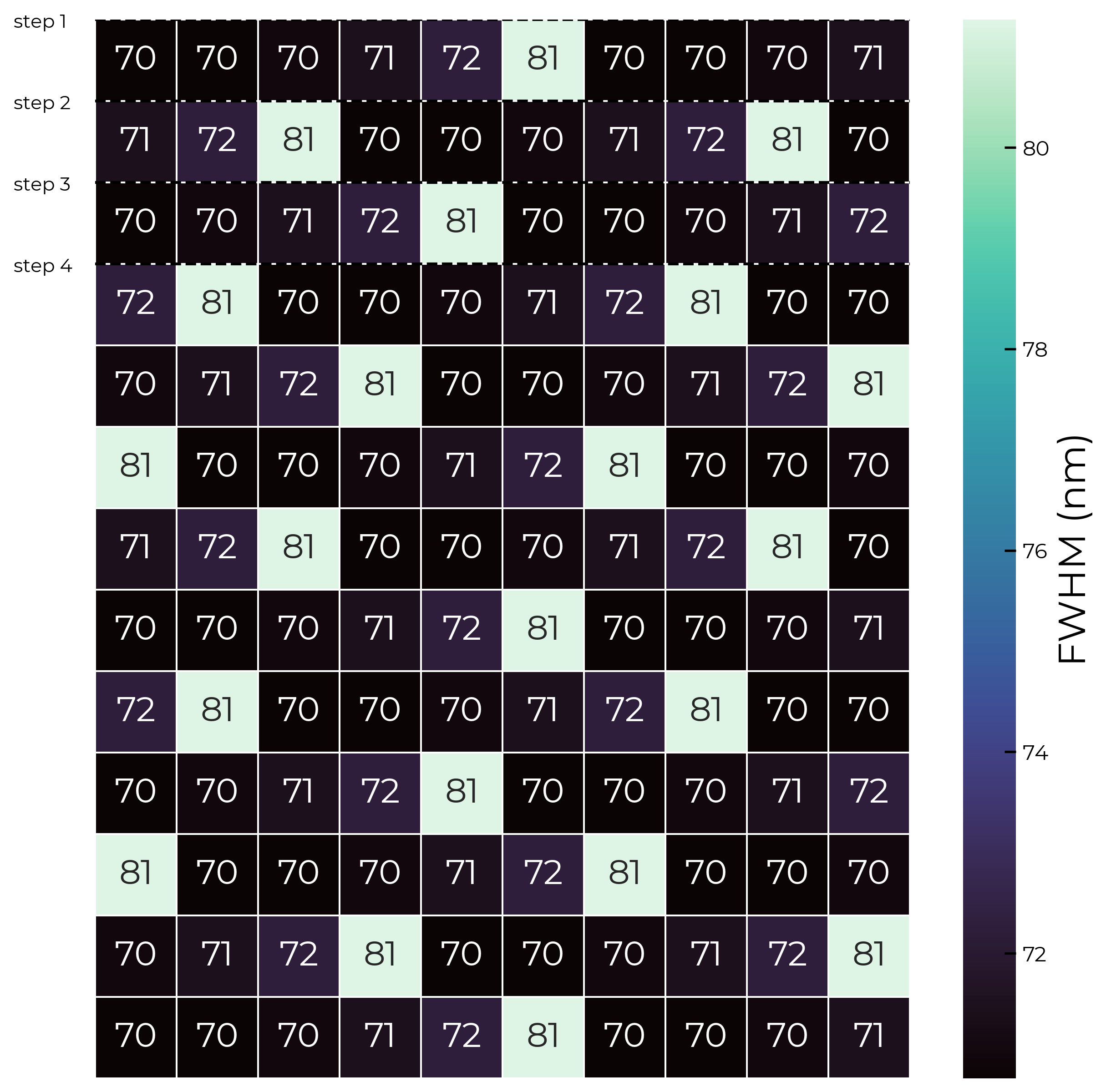}
	\caption{Squarish layout with optimized filters.}
	\label{figure: squarish joint detector}
	\end{subfigure}
	\hfill
	\begin{subfigure}[t]{0.33\textwidth}
	\includegraphics[width = \textwidth]{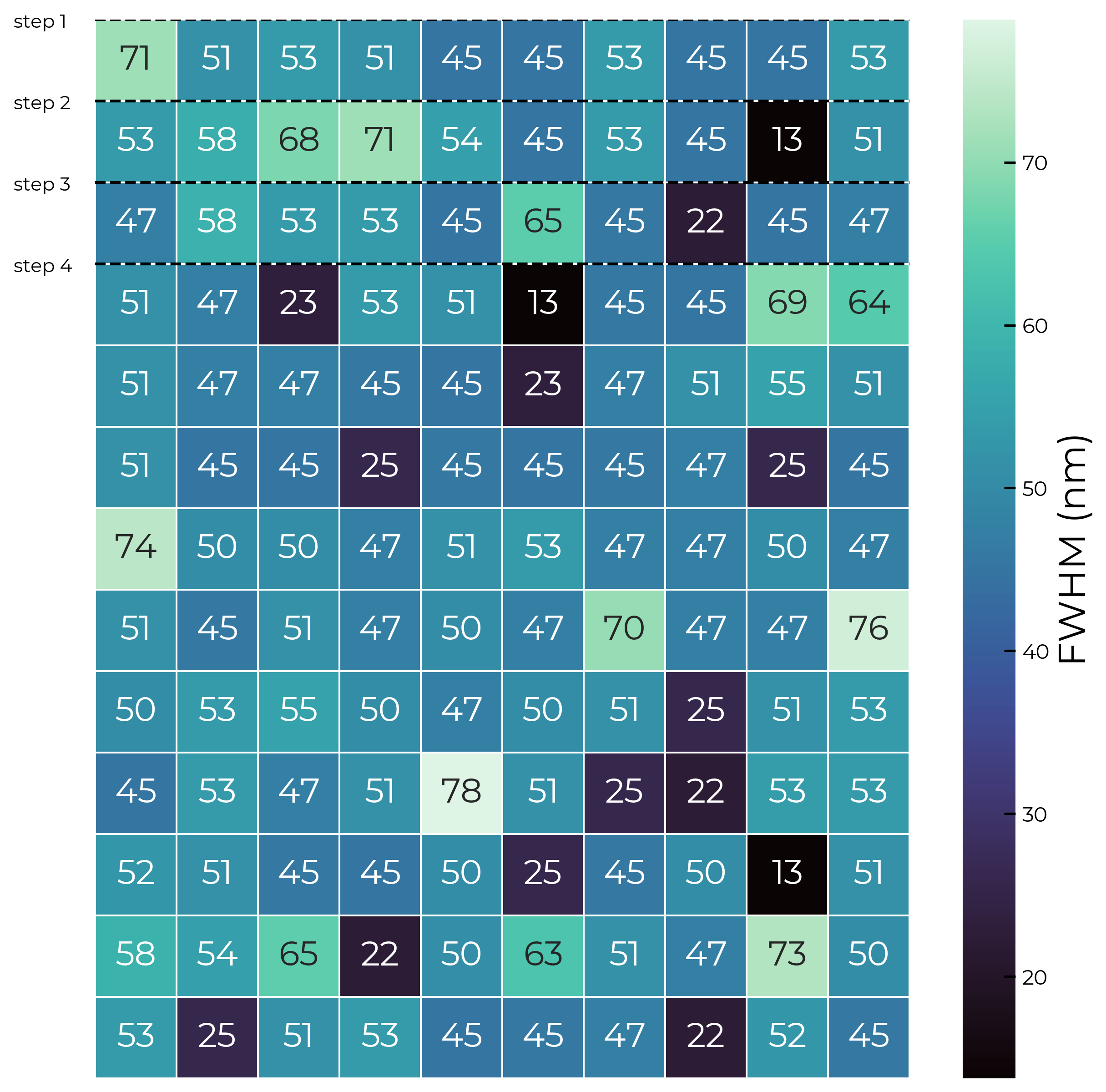}
	\caption{Optimized layout with fixed filters separately optimized by the first component.}
	\label{figure: joint detector}
	\end{subfigure}
	\caption{The detector layout in different configurations after convergence by the network. Note that the layout in \ref{figure: classical detector},  \ref{figure: bfregu detector}, and \ref{figure: squarish fixed detector} are unchanged from their initialization.}
	\label{figure: detector layouts} 
\end{figure}

We investigated five different configurations. The three best-performing configurations out of these five will be discussed in more detail. In order of complexity, the five different configurations are: 
Regular filters in a fixed LVF-like layout, Best filters in a fixed LVF-like layout, Best filters in an optimized random layout, Regular filters in a fixed "squarish" layout, and finally optimized filters in a fixed "squarish" layout. The resulting five layouts for the case of 6 filters and 4 steps are shown in figure \ref{figure: detector layouts}. 


Additional hyperparameters encode the learning rate and weight of the l2 regularization on the linear reconstructor. We run the framework with the different values of these two additional hyperparameters noted in Table \ref{Table: hyperparameters}. The resulting validation losses are then compared to determine the optimum design and reconstructor for each configuration. Furthermore, the "joint" configuration is run multiple times, starting with different random initializations of the spectral filter Layer, which is not necessary for the other configurations that feature a static spectral filter layout.

\begin{table}
\centering
\caption{The different possible values of the hyperparameters.}
\captionsetup{width=.6\linewidth}
\label{Table: hyperparameters}
\begin{tabular}{|l|l|l|l|}
\hline
 \textbf{L2 weight} &  0 & 0.0001 & 0.001  \\
 \hline
 \textbf{Learning rate} & 0.0001 & 0.0003 & 0.001  \\
 \hline
\end{tabular}
\end{table}

\begin{figure}
	\centering
	\captionsetup{width=.65\linewidth}
	\includegraphics[width=0.65\textwidth]{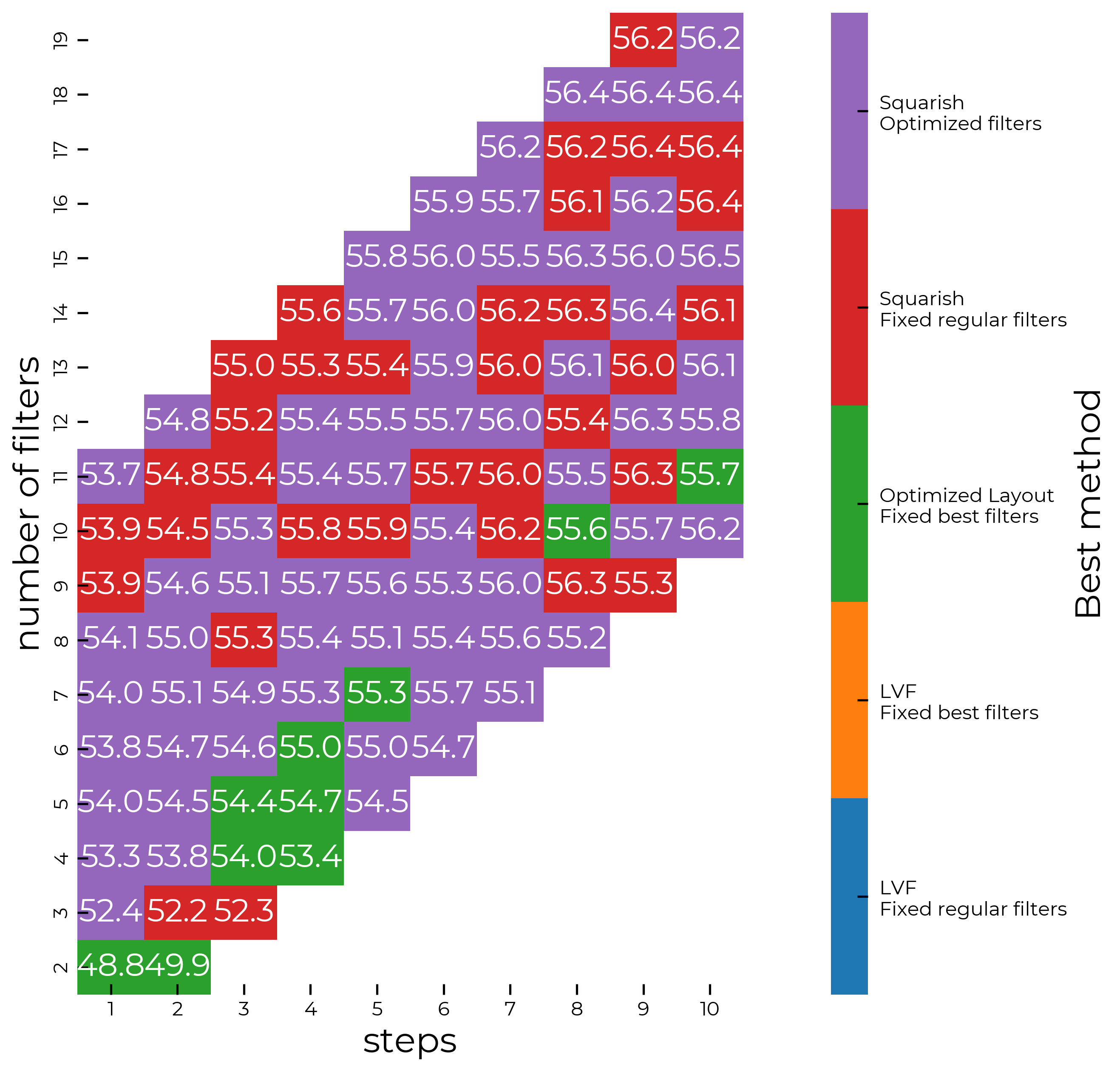}
	\caption{Color-coded map to indicate what set-up produces the most accurate results. In each block the best achievable PSNR is noted, with a color corresponding to which set-up has achieved this.}
	\label{figure: best method}
\end{figure}

Figure \ref{figure: best method} shows the configuration that reaches the best PSNR for a given pair of steps and filters. We can see that the "squarish" pattern with optimized filters performs best in almost all cases. When the number of filters is high, the difference between optimized filters and just regular filters begins to diminish.   


Every additional push broom step implies an additional image that has to be acquired and transmitted by the instrument. The original LVF design of the Hyperscout instrument needs 40 push broom steps, which requires 40 images to construct the full data cube. We define compression as the amount of images that need to be taken compared to the original 40 images. This is directly related to compression in datarate and acquistion time, since datarate is the amount of data that has to be downlinked, or the sum of images, and acquisition time is the time it takes to make all the images.  

\begin{figure}
	\centering
	\captionsetup{width=.45\linewidth}
	\includegraphics[width = 0.45\textwidth]{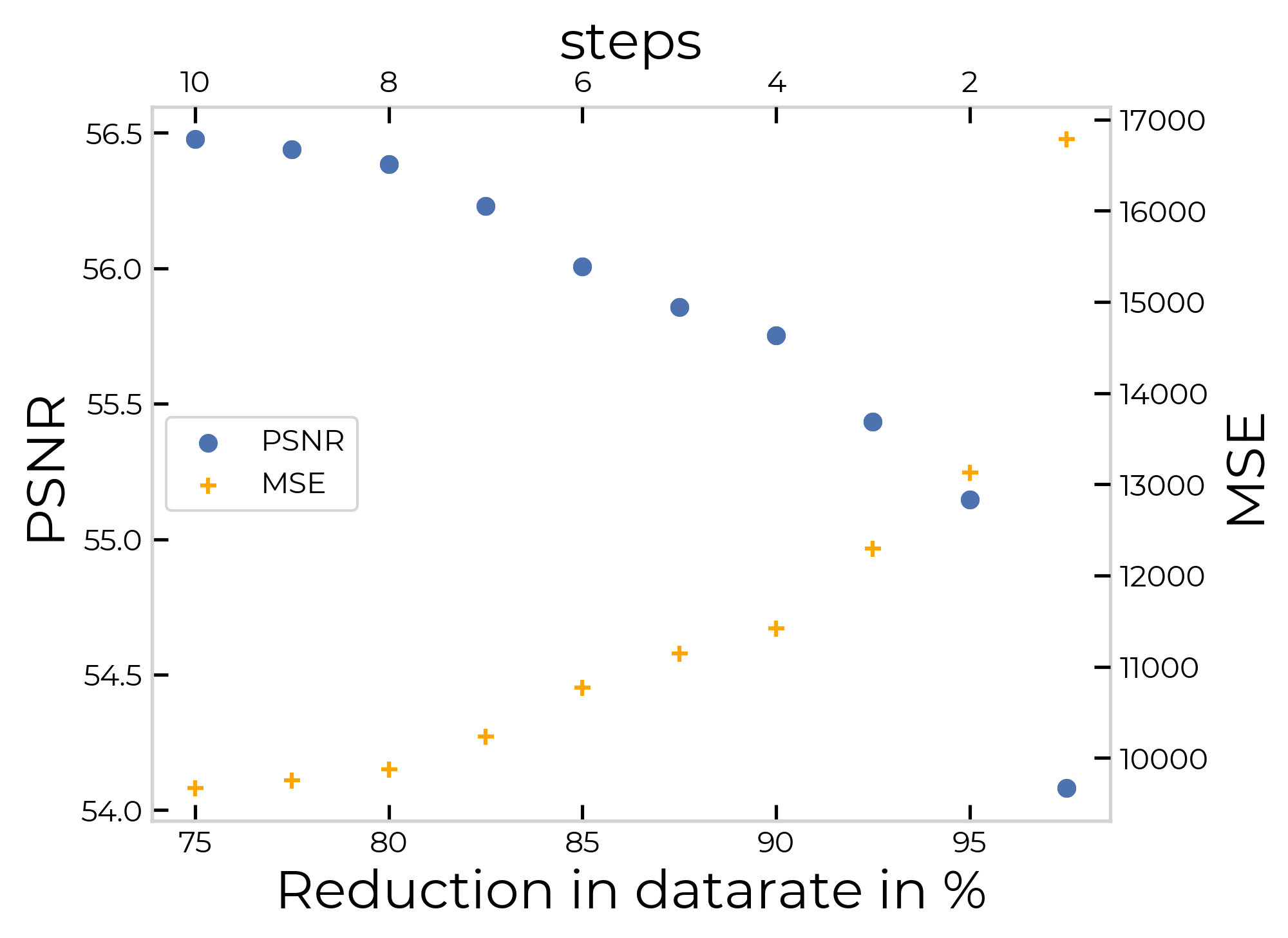}
	\caption{The reduction in data rate that can be achieved compared to the classical LVF set-up}
	\label{figure: acc_gain}
\end{figure}

Figure \ref{figure: acc_gain} shows the accuracy that can be achieved for a given fraction of the data rate of the original LVF set-up. The reduction of the data rate is only related to the number of push broom steps, not to the number of filters.  Therefore, the y-axis of this figure is the highest PSNR at each number of steps; a data rate reduction of 95\% then corresponds to two steps, i.e.\ taking only two images. In figure \ref{figure: best method} we can see that the highest possible PSNR for two push broom steps (denoted on the x-axis) occurs at seven filters (denoted on the y-axis) by the Squarish optimized filters (denoted by the color purple), which corresponds to the quoted PSNR in figure \ref{figure: acc_gain}. As expected, higher compression leads to lower accuracy. With a compression by a factor of 40 (snapshot), the achievable PSNR is still 54.1.

We could estimate the expected compression rate in chapter \ref{chapter: data exploration}, where we showed the power spectrum and the PCA results. The power spectrum showed spatial correlations. We could see that most of the power is concentrated in lower frequencies. At z>100 the power has dropped by 4 orders of magnitude. Removing all the information at the z>100 frequencies would therefore only reduce the accuracy by 1\%. This corresponds with a compression of a factor two in both spatial directions.

From the PCA analysis, we expected seven filters to be enough to recover about 99\% of the spectral information. Compared to the 40 original filters, we expected a possible compression rate of around six times in the spectral dimension.

Multiplying these expected compression rates, we expect to be able to compress the data by a factor of about 24. Figure \ref{figure: acc_gain} indeed indicates the largest drop-off in quality occurs between compression factors of 20 (95\%)to a factor of 40 (97.5\%). 

\begin{figure}
	\begin{subfigure}[t]{0.45\textwidth}
		\includegraphics[width=\textwidth]{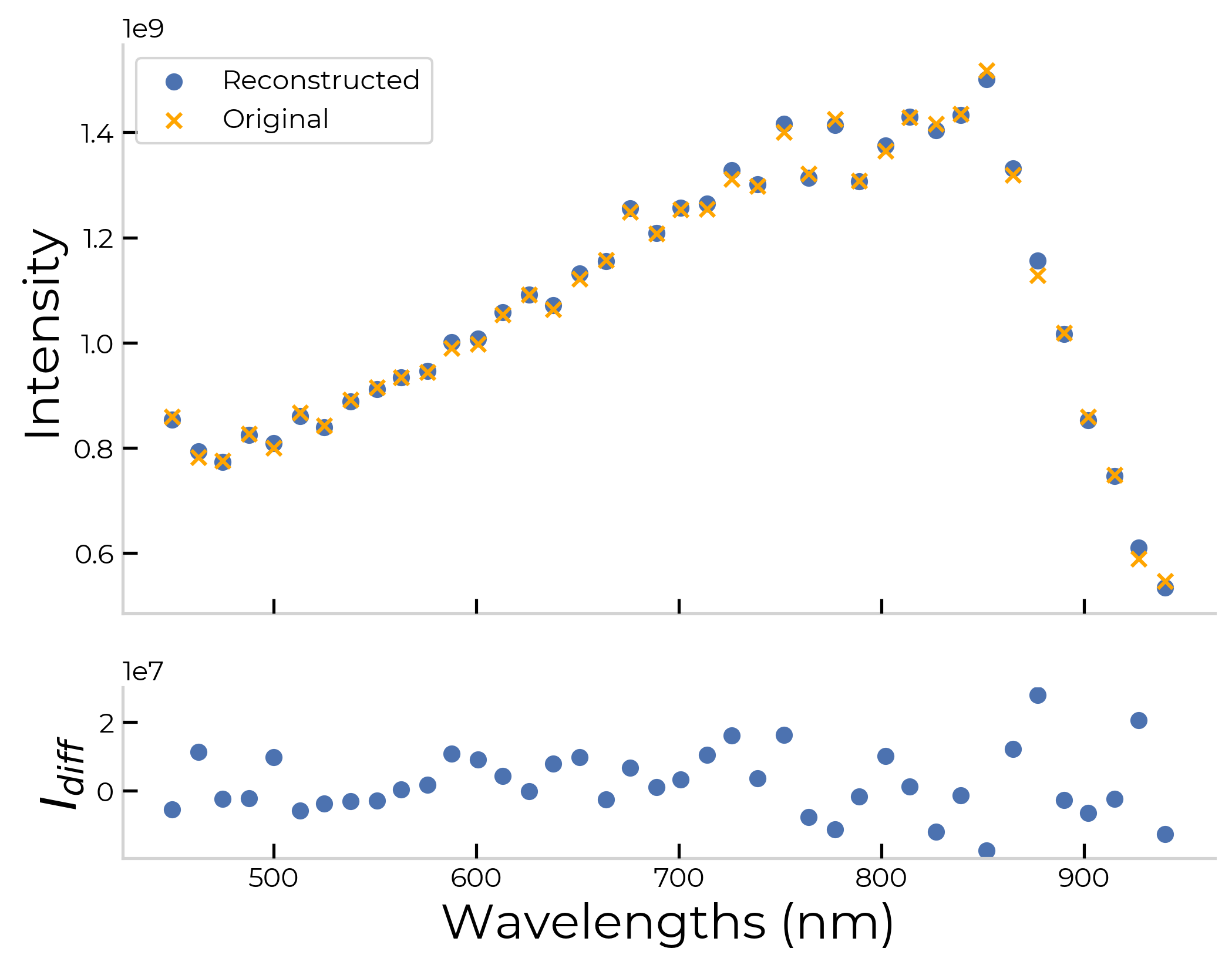}
		\caption{PSNR of 54.1}
	\end{subfigure}
	\begin{subfigure}[t]{0.45\textwidth}
		\includegraphics[width=\textwidth]{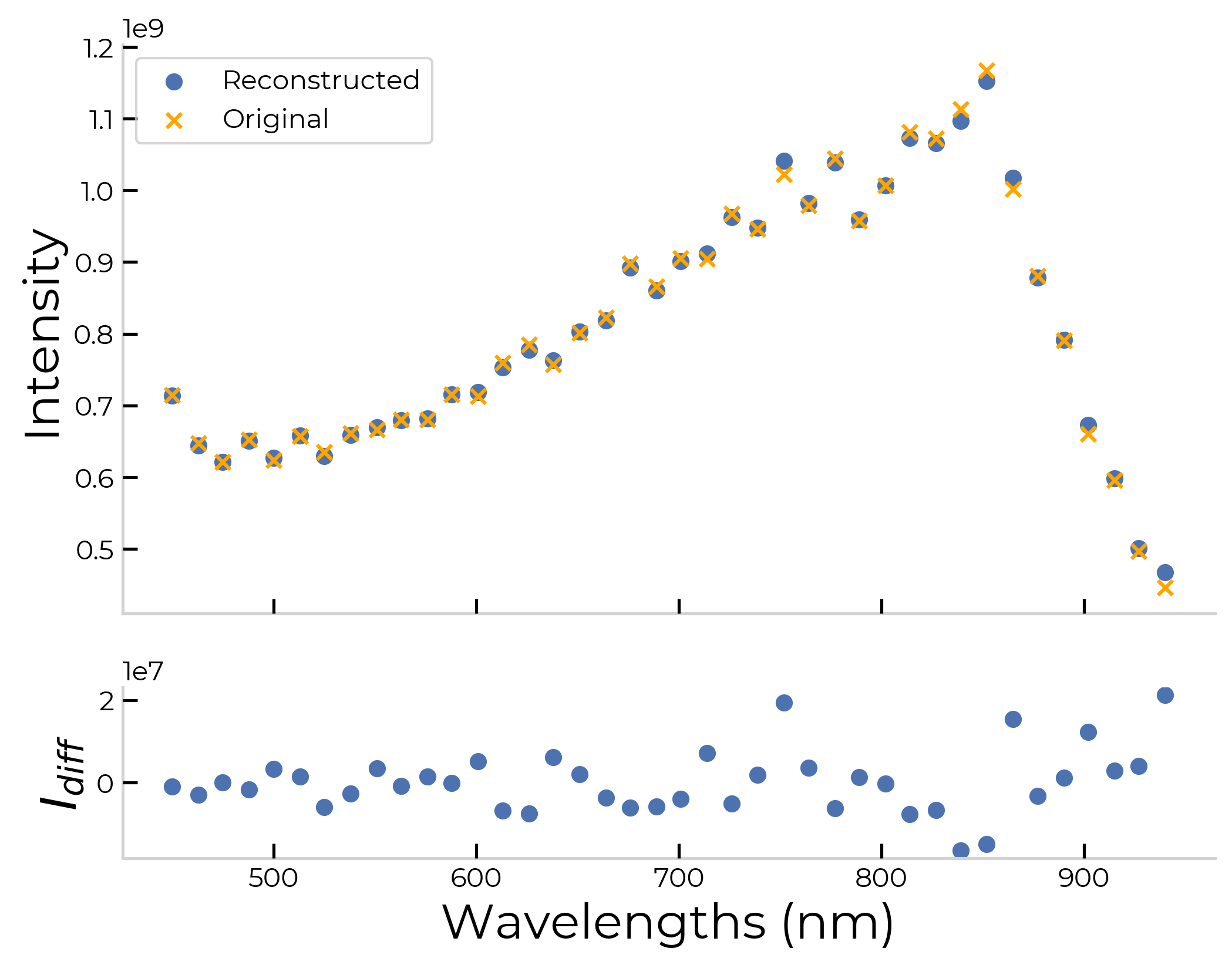}
		\caption{PSNR of 56.5}
	\end{subfigure}
    \caption{Two comparisons of the original spectrum (orange crosses) with a retrieved spectrum (blue dots) are shown for two different PSNR levels. Both spectra come from the median performing datacubes from the test set after having been thrown into the design of the leftmost and rightmost point of figure \ref{figure: acc_gain}}
    \label{figure: PSNR_feeling}
\end{figure}

\begin{figure}
	\begin{subfigure}[t]{0.45\textwidth}
		\includegraphics[width=\textwidth]{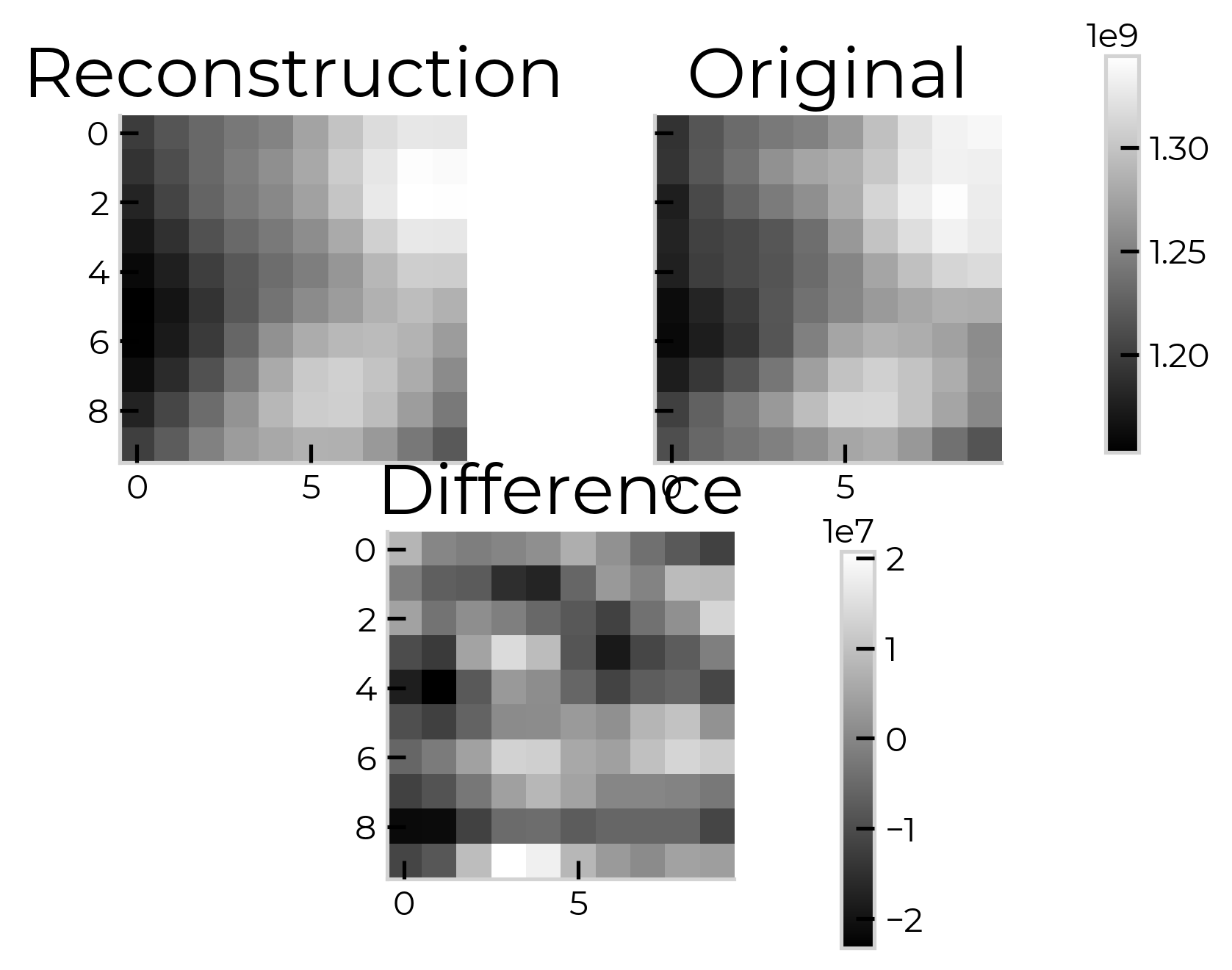}
		\caption{PSNR of 54.1}
	\end{subfigure}
	\begin{subfigure}[t]{0.45\textwidth}
		\includegraphics[width=\textwidth]{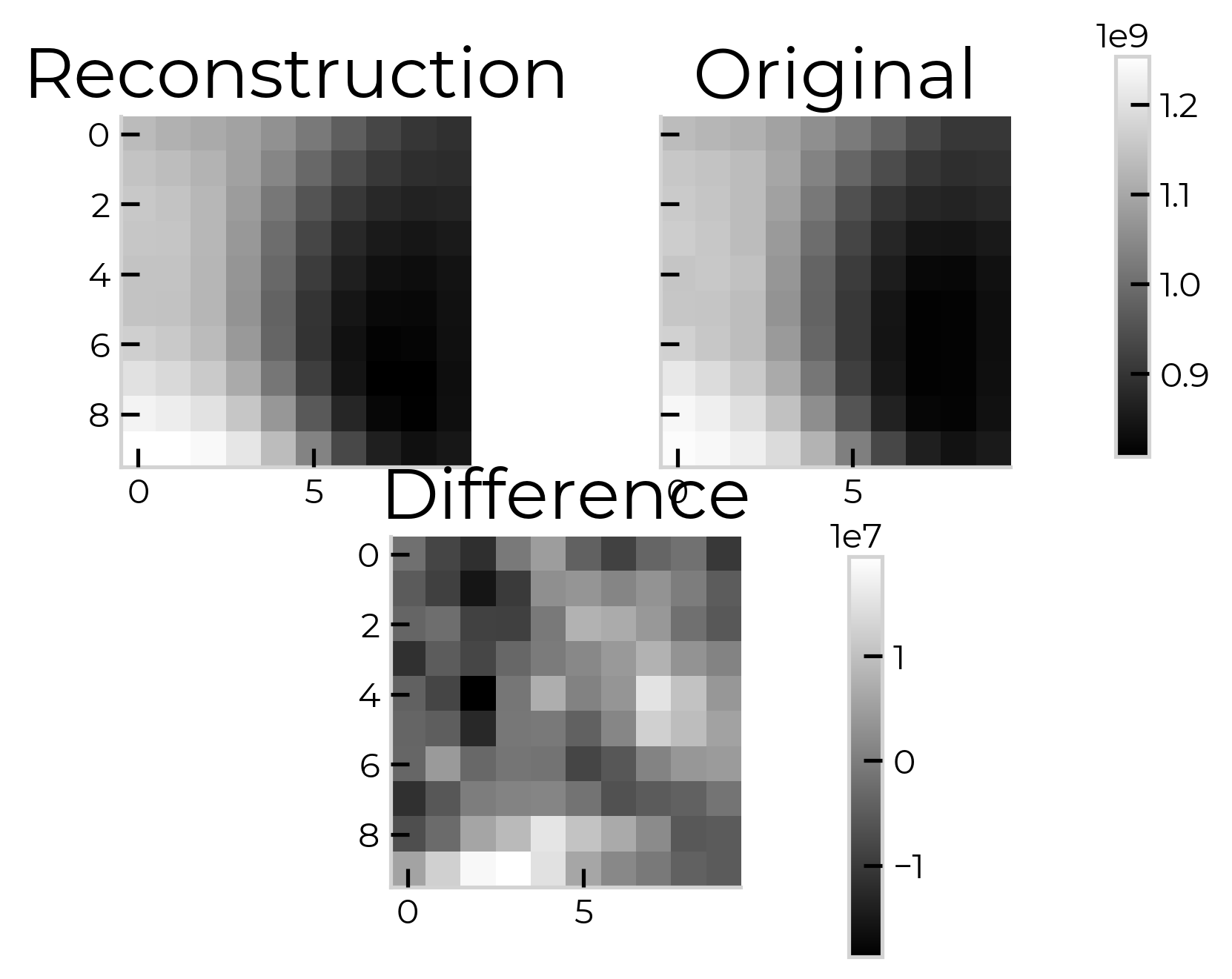}
		\caption{PSNR of 56.5}
	\end{subfigure}
    \caption{Comparison of the 701 nm intensity images, where the reconstruction is to the left and the original to the right. The difference between the two is in the figure underneath. Both images come from the median performing datacubes from the test set after having been thrown into the design of the leftmost and rightmost point of figure \ref{figure: acc_gain}}
    \label{figure: PSNR_feeling_image}
\end{figure}

In order to give a better understanding of the difference between a PSNR of 54.1 and one of 56.5, we have included Figures \ref{figure: PSNR_feeling} \& \ref{figure: PSNR_feeling_image}. In these figures, we show the difference in spectra and spatial images. At this level, the difference between how well the reconstructions are done becomes hard to discern by eye and the use of the figures of merit over visual inspection becomes apparent.   

\section{Discussion}
The inSPECtor framework designs pixelated spectral filter layouts in the focal plane along with a linear reconstructor. The resulting instruments are expected to achieve high accuracy while substantially reducing the data rate and/or acquisition time.

The results noted in this paper are a proof of concept of this framework. In the actual use of this framework it is highly advisable to make use of a more diverse training set than the single hyperspectral datacube used throughout this paper. The best results are expected to come from a training set that contains all the expected scenes in a balanced manner. 

In some cases the optimized configuration was outperformed by the equivalent non-optimized filter arrangement. This would not be expected since the static configurations are within the solution space of the optimizer. If the performance of the static cases is better than the performance of the optimized design, the latter has not converged to its optimal solution. 

However, the data-driven optimization using gradient descent is not always able to converge to the global minimum of the problem, as can be seen by the comparisons between different optimization algorithms in \cite{kingma_adam_2017}. With respect to our results, this means that the optimized design generally converges to a local minimum and that this minimum can be slightly worse than either other local minima or the global minimum. Which local minimum the network converges to depends on both the type of gradient descent algorithm and the initialization of the weights. When a static configuration outperforms the optimizable counterpart, the layout will be at or close to one of those better-performing local minima. However, these differences are no more than a PSNR difference of 0.9, or a MSE difference of a factor of 1.2.

When the amount of filters becomes large, the difference between the filters optimized by the optimal filters estimator (section \ref{section: optimal filters}) and the regularly spaced filters also becomes small, and their performance becomes similar (at a maximum difference of 0.4 in PSNR). The optimal filters from the first component were estimated without regard to the spatial information, which could influence the best choice of filters. 


What we show above is a comparison of the different results of our framework. Comparing with the results of other papers, we note a higher PSNR than previous joint design algorithms by Henz et al. \cite{henz_deep_2018} or Chakrabarti \cite{chakrabarti_learning_2016}. However, they focus on a different data product, an RGB image instead of a hyperspectral data cube. Jacome et al. \cite{jacome_d2uf_2022} look at the hyperspectral retrieval as we do, but make use of an additional CASSI instrument. We could compare our result of taking a snapshot image with 3 filters to results from spectral recovery \cite{arad_ntire_2022-1} where they start with a snapshot image made with 3 filters (RGB). However, this would only constitute a comparison of the best methods for spectral recovery on an actual camera with a basic linear reconstructor on a simulated detector. The linear reconstructor is something that can be interchanged, as will be mentioned in the future outlook section of the conclusion (chapter \ref{chapter: conclusion}), so this would not be an informative comparison.

It is important to note that the trained reconstructor for a given filter layout design is not the optimal reconstructor for the actual hardware. The reconstructor that comes with the design assumes a perfect Gaussian filter profile, a perfect match of the filters to the detector pixels, and an ideal performance of the detector at all wavelengths; all these assumptions will not hold in reality. The transmission profile for the filters will tend more towards a top-hat function, especially in the case of broader bandwidths. The match of the filters to the detector pixels is plagued by slight optical misalignment or other manufacturing errors. Finally, the detector has a wavelength-dependent efficiency and a non-zero background, which is not uniform over the whole array.   

However, as long as all the optical and electrical components are still within a linear regime, where their response is linearly related to the number of photons entering the sensor, a linear reconstructor can still be used. When the response of the Analog-to-Digital converter, for instance, does not scale linearly with the infalling intensity of the light anymore, a linear reconstructor should not be expected to return accurate approximations of the hyperspectral data cube. In the former case, the weights of the linear reconstructor, however, should still be relearned and cannot be copied from the reconstructor that came with the design. The relearning is done by feeding intensity images of calibrated/known sources created by the prototype into the linear reconstructor part of the optimal filter layout estimator and optimizing the weights for the reconstruction.  

Finally, the goal of acquiring hyperspectral data goes beyond the acquisition of the data cube itself; instead, the final data product often requires further post-processing like segmentation and classification \cite{audebert_deep_2019, imani_overview_2020, rasti_feature_2020}. One of the strengths of our design tool is that these post-processing steps can be implemented right before the calculation of the loss function if they can be described in a differentiable manner. The loss function should then be modified to reflect the quality of the results after post-processing. 

\section{Conclusion}
\label{chapter: conclusion}
In this paper, we describe a new tool for the design of spectral filter layouts on pixelated detectors as used for compressed, hyperspectral imaging. The resulting filter layout makes a partial measurement of the spectrum of every pixel. However, due to the simultaneously optimized linear reconstructor, this partial measurement can be used to reconstruct the full hyperspectral data cube with high accuracy. We show that the network can converge towards a filter layout that can recover known scenes with a snapshot image and high accuracy. This opens possibilities for extremely compact hyperspectral imagers with low data rates and short acquisition times.  

As of now, InSPECtor does not yet do a full joint optimization since the calculation of the optimal filters is still separate from the calculation of the optimal layout. This could be changed in future adaptations. Other possible additions include making the functional form of the spectral filters variable, so it can diverge from the Gaussian form it has now. A major change, carrying potentially big benefits, would be to replace the linear reconstructor with a non-linear algorithm such as a neural network or one of the algorithms mentioned in the introduction.   

InSPECtor can be used to design a multitude of filter layouts for pixelated, hyperspectral imagers. The algorithm presented in this paper is easily adaptable to different scenes and applications since it is a matter of optimizing InSPECtor with data comparable to the scene of interest. These scenes can range from remote sensing, to astronomy or defect inspection in factories. The framework could also be adapted to go further than hyperspectral imagers. Additional optical components, e.g.\ an array of linear polarizers \cite{stockmans_end--end_2022}, can be added before or after the existing Layers if they can be mathematically described in a differentiable and continuous manner. Finally, given known design constraints such as the desired data rate, minimally desired accuracy, and filter manufacturing constraints, the framework can return the optimal design of the filter layout.

\textbf{\large{Funding.}} NWO-TTW SYNOPTICS program.

\textbf{\large{Acknowledgements.}} This work was performed using the compute resources from the Academic Leiden Interdisciplinary Cluster Environment (ALICE) provided by Leiden University.

\textbf{\large{Disclosures.}} The authors declare that there are no conflicts of interest related to this paper.

\textbf{\large{Data availability.}} Data and code underlying the results presented in this paper are not publicly available at this time but may be obtained from the authors upon reasonable request. 


\bibliography{interpol_paper_23_mod}

\begin{thebibliography}{10}
\newcommand{\enquote}[1]{``#1''}

\bibitem{bioucas-dias_hyperspectral_2012}
J.~M. Bioucas-Dias, A.~Plaza, N.~Dobigeon, M.~Parente, Q.~Du, P.~Gader, and
  J.~Chanussot, \enquote{Hyperspectral unmixing overview: Geometrical,
  statistical, and sparse regression-based approaches,}
  {\protect\JournalTitle{{IEEE} Journal of Selected Topics in Applied Earth
  Observations and Remote Sensing}} \textbf{5}, 354--379 (2012).

\bibitem{vane_airborne_1993}
G.~Vane, R.~O. Green, T.~G. Chrien, H.~T. Enmark, E.~G. Hansen, and W.~M.
  Porter, \enquote{The airborne visible/infrared imaging spectrometer
  ({AVIRIS}),} {\protect\JournalTitle{Remote Sensing of Environment}}
  \textbf{44}, 127--143 (1993).

\bibitem{barnes_status_2003}
W.~L. Barnes, X.~Xiong, and V.~V. Salomonson, \enquote{Status of terra {MODIS}
  and aqua modis,} {\protect\JournalTitle{Advances in Space Research}}
  \textbf{32}, 2099--2106 (2003).

\bibitem{elmasry_chapter_2010}
G.~{ElMasry} and D.-W. Sun, \enquote{{CHAPTER} 1 - principles of hyperspectral
  imaging technology,} in \emph{Hyperspectral Imaging for Food Quality Analysis
  and Control,}  D.-W. Sun, ed. (Academic Press, 2010), pp. 3--43.

\bibitem{gowen_hyperspectral_2007}
A.~A. Gowen, C.~P. O'Donnell, P.~J. Cullen, G.~Downey, and J.~M. Frias,
  \enquote{Hyperspectral imaging – an emerging process analytical tool for
  food quality and safety control,} {\protect\JournalTitle{Trends in Food
  Science \& Technology}} \textbf{18}, 590--598 (2007).

\bibitem{liang_advances_2012}
H.~Liang, \enquote{Advances in multispectral and hyperspectral imaging for
  archaeology and art conservation,} {\protect\JournalTitle{Appl. Phys. A}}
  \textbf{106}, 309--323 (2012).

\bibitem{hege_hyperspectral_2004}
E.~K. Hege, D.~O'Connell, W.~Johnson, S.~Basty, and E.~L. Dereniak,
  \enquote{Hyperspectral imaging for astronomy and space surviellance,} in
  \emph{Optical Science and Technology, {SPIE}'s 48th Annual Meeting,}  S.~S.
  Shen and P.~E. Lewis, eds. (2004), p. 380.

\bibitem{adao_hyperspectral_2017}
T.~Adão, J.~Hruška, L.~Pádua, J.~Bessa, E.~Peres, R.~Morais, and J.~J.
  Sousa, \enquote{Hyperspectral imaging: A review on {UAV}-based sensors, data
  processing and applications for agriculture and forestry,}
  {\protect\JournalTitle{Remote Sensing}} \textbf{9}, 1110 (2017). Number: 11
  Publisher: Multidisciplinary Digital Publishing Institute.

\bibitem{fei_chapter_2020}
B.~Fei, \enquote{Chapter 3.6 - hyperspectral imaging in medical applications,}
  in \emph{Data Handling in Science and Technology,}  vol.~32 of
  \emph{Hyperspectral Imaging} J.~M. Amigo, ed. (Elsevier, 2020), pp. 523--565.

\bibitem{ortega_hyperspectral_2020}
S.~Ortega, M.~Halicek, H.~Fabelo, G.~M. Callico, and B.~Fei,
  \enquote{Hyperspectral and multispectral imaging in digital and computational
  pathology: a systematic review [invited],} {\protect\JournalTitle{Biomed.
  Opt. Express}} \textbf{11}, 3195 (2020).

\bibitem{dong_dmd-based_2021}
X.~Dong, G.~Tong, X.~Song, X.~Xiao, and Y.~Yu, \enquote{{DMD}-based
  hyperspectral microscopy with flexible multiline parallel scanning,}
  {\protect\JournalTitle{Microsyst Nanoeng}} \textbf{7}, 68 (2021).

\bibitem{eismann_hyperspectral_2012}
M.~T. Eismann, \emph{Hyperspectral remote sensing} ({SPIE} Press, 2012).
  {OCLC}: ocn768041864.

\bibitem{li_review_2013}
Q.~Li, X.~He, Y.~Wang, H.~Liu, D.~Xu, and F.~Guo, \enquote{Review of spectral
  imaging technology in biomedical engineering: achievements and challenges,}
  {\protect\JournalTitle{J. Biomed. Opt}} \textbf{18}, 100901 (2013).

\bibitem{ustin_current_2021}
S.~L. Ustin and E.~M. Middleton, \enquote{Current and near-term advances in
  earth observation for ecological applications,} {\protect\JournalTitle{Ecol
  Process}} \textbf{10}, 1 (2021).

\bibitem{willett_sparsity_2014}
R.~M. Willett, M.~F. Duarte, M.~A. Davenport, and R.~G. Baraniuk,
  \enquote{Sparsity and structure in hyperspectral imaging : Sensing,
  reconstruction, and target detection,} {\protect\JournalTitle{{IEEE} Signal
  Processing Magazine}} \textbf{31}, 116--126 (2014).

\bibitem{guzzi_donatella_optical_2019}
{Guzzi, Donatella}, G.~Coluccia, D.~Labate, C.~Lastri, E.~Magli, V.~Nardino,
  L.~Palombi, I.~Pippi, D.~Coltuc, A.~Z. Marchi, and V.~Raimondi,
  \enquote{Optical compressive sensing technologies for space applications:
  instrumental concepts and performance analysis,} in \emph{International
  Conference on Space Optics — {ICSO} 2018,}  (2019). {MAG} {ID}: 2921500050.

\bibitem{sun_hyperspectral_2019}
W.~Sun and Q.~Du, \enquote{Hyperspectral band selection: A review,}
  {\protect\JournalTitle{{IEEE} Geosci. Remote Sens. Mag.}} \textbf{7},
  118--139 (2019).

\bibitem{cao_computational_2016}
X.~Cao, T.~Yue, X.~Lin, S.~Lin, X.~Yuan, Q.~Dai, L.~Carin, and D.~J. Brady,
  \enquote{Computational snapshot multispectral cameras: Toward dynamic capture
  of the spectral world,} {\protect\JournalTitle{{IEEE} Signal Process. Mag.}}
  \textbf{33}, 95--108 (2016).

\bibitem{coluccia_optical_2020}
G.~Coluccia, C.~Lastri, D.~Guzzi, E.~Magli, V.~Nardino, L.~Palombi, I.~Pippi,
  V.~Raimondi, C.~Ravazzi, F.~Garoi, D.~Coltuc, R.~Vitulli, and A.~Z. Marchi,
  \enquote{Optical compressive imaging technologies for space big data,}
  {\protect\JournalTitle{{IEEE} Trans. Big Data}} \textbf{6}, 430--442 (2020).

\bibitem{barducci_compressive_2014}
A.~Barducci, D.~Guzzi, C.~Lastri, V.~Nardino, I.~Pippi, and V.~Raimondi,
  \enquote{Compressive sensing for hyperspectral earth observation from space,}
  in \emph{International Conference on Space Optics — {ICSO} 2014,}  vol.
  10563 B.~Cugny, Z.~Sodnik, and N.~Karafolas, eds. ({SPIE}, 2014), pp. 18--18.
  Issue: October.

\bibitem{okamoto_simultaneous_1991}
T.~Okamoto and I.~Yamaguchi, \enquote{Simultaneous acquisition of spectral
  image information,} {\protect\JournalTitle{Opt. Lett.}} \textbf{16}, 1277
  (1991).

\bibitem{wagadarikar_single_2008}
A.~Wagadarikar, R.~John, R.~Willett, and D.~Brady, \enquote{Single disperser
  design for coded aperture snapshot spectral imaging,}
  {\protect\JournalTitle{Applied Optics}} \textbf{47} (2008).

\bibitem{arce_compressive_2014}
G.~R. Arce, D.~J. Brady, L.~Carin, H.~Arguello, and D.~S. Kittle,
  \enquote{Compressive coded aperture spectral imaging: An introduction,}
  {\protect\JournalTitle{{IEEE} Signal Processing Magazine}} \textbf{31},
  105--115 (2014).

\bibitem{gehm_single-shot_2007}
M.~E. Gehm, R.~John, D.~J. Brady, R.~M. Willett, and T.~J. Schulz,
  \enquote{Single-shot compressive spectral imaging with a dual-disperser
  architecture,} {\protect\JournalTitle{Optics Express}} \textbf{15},
  14013--14013 (2007).

\bibitem{wu_development_2011}
Y.~Wu, I.~O. Mirza, G.~R. Arce, and D.~W. Prather, \enquote{Development of a
  digital-micromirror-device-based multishot snapshot spectral imaging system,}
  {\protect\JournalTitle{Opt. Lett., {OL}}} \textbf{36}, 2692--2694 (2011).
  Publisher: Optical Society of America.

\bibitem{august_compressive_2013}
Y.~August, C.~Vachman, Y.~Rivenson, and A.~Stern, \enquote{Compressive
  hyperspectral imaging by random separable projections in both the spatial and
  the spectral domains,} {\protect\JournalTitle{Applied Optics}} \textbf{52}
  (2013).

\bibitem{kar_compressive_2019}
O.~F. Kar and F.~S. Oktem, \enquote{Compressive spectral imaging with
  diffractive lenses,} {\protect\JournalTitle{Opt. Lett., {OL}}} \textbf{44},
  4582--4585 (2019). Publisher: Optical Society of America.

\bibitem{monakhova_spectral_2020}
K.~Monakhova, K.~Yanny, N.~Aggarwal, and L.~Waller, \enquote{Spectral
  {DiffuserCam}: lensless snapshot hyperspectral imaging with a spectral filter
  array,} {\protect\JournalTitle{Optica}} \textbf{7} (2020).

\bibitem{jin_hyperspectral_2017}
S.~Jin, W.~Hui, Y.~Wang, K.~Huang, Q.~Shi, C.~Ying, D.~Liu, Q.~Ye, W.~Zhou, and
  J.~Tian, \enquote{Hyperspectral imaging using the single-pixel fourier
  transform technique,} {\protect\JournalTitle{Sci Rep}} \textbf{7}, 45209
  (2017).

\bibitem{tropp_computational_2010}
J.~A. Tropp and S.~J. Wright, \enquote{Computational methods for sparse
  solution of linear inverse problems,} {\protect\JournalTitle{Proc. {IEEE}}}
  \textbf{98}, 948--958 (2010).

\bibitem{wang_compressed_2015}
L.~Wang, K.~Lu, and P.~Liu, \enquote{Compressed sensing of a remote sensing
  image based on the priors of the reference image,}
  {\protect\JournalTitle{{IEEE} Geosci. Remote Sensing Lett.}} \textbf{12},
  736--740 (2015).

\bibitem{yang_hyperspectral_2021}
Y.~Yang, Y.~Xie, X.~Chen, and Y.~Sun, \enquote{Hyperspectral snapshot
  compressive imaging with non-local spatial-spectral residual network,}
  {\protect\JournalTitle{Remote Sensing}} \textbf{13}, 1812 (2021).

\bibitem{gozcu_learning-based_2018}
B.~Gözcü, R.~K. Mahabadi, Y.-H. Li, E.~Ilıcak, T.~Çukur, J.~Scarlett, and
  V.~Cevher, \enquote{Learning-based compressive {MRI},}
  {\protect\JournalTitle{{IEEE} Transactions on Medical Imaging}} pp. 1394 --
  1406 (2018).

\bibitem{wu_learning_2019}
S.~Wu, A.~Dimakis, S.~Sanghavi, F.~Yu, D.~Holtmann-Rice, D.~Storcheus,
  A.~Rostamizadeh, and S.~Kumar, \enquote{Learning a compressed sensing
  measurement matrix via gradient unrolling,} in \emph{Proceedings of the 36th
  International Conference on Machine Learning,}  ({PMLR}, 2019), pp.
  6828--6839. {ISSN}: 2640-3498.

\bibitem{li_learning_2016}
Y.-H. Li and V.~Cevher, \enquote{Learning data triage: Linear decoding works
  for compressive {MRI},} in \emph{2016 {IEEE} International Conference on
  Acoustics, Speech and Signal Processing ({ICASSP}),}  (2016), pp. 4034--4038.
  {ISSN}: 2379-190X.

\bibitem{baldassarre_learning-based_2016}
L.~Baldassarre, Y.-H. Li, J.~Scarlett, B.~Gözcü, I.~Bogunovic, and V.~Cevher,
  \enquote{Learning-based compressive subsampling,}
  {\protect\JournalTitle{{IEEE} J. Sel. Top. Signal Process.}} \textbf{10},
  809--822 (2016).

\bibitem{mait_computational_2018}
J.~N. Mait, G.~W. Euliss, and R.~A. Athale, \enquote{Computational imaging,}
  {\protect\JournalTitle{Adv. Opt. Photon.}} \textbf{10}, 409 (2018).

\bibitem{gao_computational_2022}
L.~Gao, Y.~Qu, L.~Wang, and Z.~Yu, \enquote{Computational spectrometers enabled
  by nanophotonics and deep learning,} {\protect\JournalTitle{Nanophotonics}}
  \textbf{11}, 2507--2529 (2022).

\bibitem{arguello_deep_2022}
H.~Arguello, J.~Bacca, H.~Kariyawasam, E.~Vargas, M.~Marquez, R.~Hettiarachchi,
  H.~Garcia, K.~Herath, U.~Haputhanthri, B.~S. Ahluwalia, P.~So, D.~N.
  Wadduwage, and C.~U.~S. Edussooriya, \enquote{Deep optical coding design in
  computational imaging,}  (2022).

\bibitem{huang_spectral_2022}
L.~Huang, R.~Luo, X.~Liu, and X.~Hao, \enquote{Spectral imaging with deep
  learning,} {\protect\JournalTitle{Light Sci Appl}} \textbf{11}, 61 (2022).

\bibitem{bacca_computational_2023}
J.~Bacca, E.~Martinez, and H.~Arguello, \enquote{Computational spectral
  imaging: a contemporary overview,} {\protect\JournalTitle{J. Opt. Soc. Am. A,
  {JOSAA}}} \textbf{40}, C115--C125 (2023). Publisher: Optica Publishing Group.

\bibitem{wang_hyperreconnet_2019}
L.~Wang, T.~Zhang, Y.~Fu, and H.~Huang, \enquote{{HyperReconNet}: Joint coded
  aperture optimization and image reconstruction for compressive hyperspectral
  imaging,} {\protect\JournalTitle{{IEEE} Transactions on Image Processing}}
  \textbf{28}, 2257--2270 (2019). Publisher: {IEEE}.

\bibitem{kaur_survey_2015}
E.~S. Kaur and V.~K. Banga, \enquote{A survey of demosaicing : Issues and
  challenges,} {\protect\JournalTitle{International Journal of Science,
  Engineering and Technologies ({IJSET})}} \textbf{2}, 9--17 (2015).

\bibitem{bayer_color_1976}
B.~E. Bayer, \enquote{Color imaging array patent {US} 3971065,}
  {\protect\JournalTitle{U. S. patent 3971065}} pp. 10--10 (1976).

\bibitem{gharbi_deep_2016}
M.~Gharbi, G.~Chaurasia, S.~Paris, and F.~Durand, \enquote{Deep joint
  demosaicking and denoising,} {\protect\JournalTitle{{ACM} Transactions on
  Graphics}} \textbf{35}, 1--12 (2016).

\bibitem{cui_color_2018}
K.~Cui, Z.~Jin, and E.~Steinbach, \enquote{Color image demosaicking using a
  3-stage convolutional neural network structure,} in \emph{2018 25th {IEEE}
  International Conference on Image Processing ({ICIP}),}  ({IEEE}, 2018), pp.
  2177--2181.

\bibitem{guo_joint_2021}
S.~Guo, Z.~Liang, and L.~Zhang, \enquote{Joint denoising and demosaicking with
  green channel prior for real-world burst images,}
  {\protect\JournalTitle{{IEEE} Transactions on Image Processing}} \textbf{30},
  6930--6942 (2021).

\bibitem{he_self-learning_2012}
F.-L. He, Y.-C.~F. Wang, and K.-L. Hua, \enquote{Self-learning approach to
  color demosaicking via support vector regression,} in \emph{2012 19th {IEEE}
  International Conference on Image Processing,}  (2012), pp. 2765--2768.

\bibitem{heinze_joint_2012}
T.~Heinze, M.~von Löwis, and A.~Polze, \enquote{Joint multi-frame demosaicing
  and super-resolution with artificial neural networks,} in \emph{2012 19th
  International Conference on Systems, Signals and Image Processing
  ({IWSSIP}),}  (2012), pp. 540--543.

\bibitem{iriyama_deep_2021}
T.~Iriyama, M.~Sato, H.~Aomori, and T.~Otake, \enquote{Deep demosaicking
  considering inter-channel correlation and self-similarity,}
  {\protect\JournalTitle{Nonlinear Theory and Its Applications, {IEICE}}}
  \textbf{12}, 453--463 (2021).

\bibitem{jin_review_2020}
Q.~Jin, G.~Facciolo, and J.~M. Morel, \enquote{A review of an old dilemma:
  Demosaicking first, or denoising first?} {\protect\JournalTitle{{IEEE}
  Computer Society Conference on Computer Vision and Pattern Recognition
  Workshops}} \textbf{2020-June}, 2169--2179 (2020).

\bibitem{kiku_beyond_2016}
D.~Kiku, Y.~Monno, M.~Tanaka, and M.~Okutomi, \enquote{Beyond color difference:
  Residual interpolation for color image demosaicking.}
  {\protect\JournalTitle{{IEEE} transactions on image processing : a
  publication of the {IEEE} Signal Processing Society}} \textbf{25}, 1288--1300
  (2016).

\bibitem{menon_color_2011}
D.~Menon and G.~Calvagno, \enquote{Color image demosaicking: An overview,}
  {\protect\JournalTitle{Signal Processing: Image Communication}} \textbf{26},
  518--533 (2011). Publisher: Elsevier.

\bibitem{sharif_beyond_2021}
S.~M.~A. Sharif, R.~Ali~Naqvi, and M.~Biswas, \enquote{Beyond joint
  demosaicking and denoising: An image processing pipeline for a pixel-bin
  image sensor,} in \emph{2021 {IEEE}/{CVF} Conference on Computer Vision and
  Pattern Recognition Workshops ({CVPRW}),}  ({IEEE}, 2021), pp. 233--242.

\bibitem{wang_multilayer_2014}
Y.-Q. Wang, \enquote{A multilayer neural network for image demosaicking,} in
  \emph{2014 {IEEE} International Conference on Image Processing ({ICIP}),}
  (2014), pp. 1852--1856.

\bibitem{wu_color_2011}
X.~Wu, \enquote{Color demosaicking by local directional interpolation
  and nonlocal adaptive thresholding,} {\protect\JournalTitle{Journal of
  Electronic Imaging}} \textbf{20}, 023016--023016 (2011).

\bibitem{zhang_learning_2018}
J.~Zhang, J.~Shao, H.~Luo, X.~Zhang, B.~Hui, Z.~Chang, and R.~Liang,
  \enquote{Learning a convolutional demosaicing network for microgrid
  polarimeter imagery,} {\protect\JournalTitle{Optics Letters}} \textbf{43},
  4534--4534 (2018).

\bibitem{habtegebrial_deep_2019}
T.~A. Habtegebrial, G.~Reis, and D.~Stricker, \enquote{Deep convolutional
  networks for snapshot hypercpectral demosaicking,} in \emph{2019 10th
  Workshop on Hyperspectral Imaging and Signal Processing: Evolution in Remote
  Sensing ({WHISPERS}),}  (2019), pp. 1--5.

\bibitem{dijkstra_hyperspectral_2019}
K.~Dijkstra, J.~van~de Loosdrecht, L.~R. Schomaker, and M.~A. Wiering,
  \enquote{Hyperspectral demosaicking and crosstalk correction using deep
  learning,} {\protect\JournalTitle{Machine Vision and Applications}}
  \textbf{30}, 1--21 (2019). Publisher: Springer Berlin Heidelberg.

\bibitem{li_deep_2021}
P.~Li, M.~Ebner, P.~Noonan, C.~Horgan, A.~Bahl, S.~Ourselin, J.~Shapey, and
  T.~Vercauteren, \enquote{Deep learning approach for hyperspectral image
  demosaicking, spectral correction and high-resolution {RGB} reconstruction,}
  {\protect\JournalTitle{{MICCAI} Workshop on Augmented Environments for
  Computer-Assisted Interventions, Computer Assisted and Robotic Endoscopy, and
  Context Aware Operating Theaters (In Print)}} pp. 12--12 (2021).

\bibitem{wang_discrete_2013}
X.~Wang, J.-B. Thomas, J.~Y. Hardeberg, and P.~Gouton, \enquote{Discrete
  wavelet transform based multispectral filter array demosaicking,} in
  \emph{2013 Colour and Visual Computing Symposium ({CVCS}),}  (2013), pp.
  1--6.

\bibitem{zhuang_hy-demosaicing_2022}
L.~Zhuang, M.~K. Ng, X.~Fu, and J.~M. Bioucas-Dias, \enquote{Hy-demosaicing:
  Hyperspectral blind reconstruction from spectral subsampling,}
  {\protect\JournalTitle{{IEEE} Trans. Geosci. Remote Sensing}} \textbf{60},
  1--15 (2022).

\bibitem{tsagkatakis_graph_2019}
G.~Tsagkatakis, M.~Bloemen, B.~Geelen, M.~Jayapala, and P.~Tsakalides,
  \enquote{Graph and rank regularized matrix recovery for snapshot spectral
  image demosaicing,} {\protect\JournalTitle{{IEEE} Trans. Comput. Imaging}}
  \textbf{5}, 301--316 (2019).

\bibitem{mihoubi_multispectral_2017}
S.~Mihoubi, O.~Losson, B.~Mathon, and L.~Macaire, \enquote{Multispectral
  demosaicing using pseudo-panchromatic image,} {\protect\JournalTitle{{IEEE}
  Trans. Comput. Imaging}} \textbf{3}, 982--995 (2017).

\bibitem{amba_n-lmmse_2017}
P.~Amba, J.~B. Thomas, and D.~Alleysson, \enquote{N-{LMMSE} demosaicing for
  spectral filter arrays,} {\protect\JournalTitle{jist}} \textbf{61},
  40407--1--40407--11 (2017).

\bibitem{arad_ntire_2022}
B.~Arad, R.~Timofte, R.~Yahel, N.~Morag, A.~Bernat, Y.~Wu, X.~Wu, Z.~Fan,
  C.~Xia, F.~Zhang, S.~Liu, Y.~Li, C.~Feng, L.~Lei, M.~Zhang, K.~Feng,
  X.~Zhang, J.~Yao, Y.~Zhao, S.~Ma, F.~He, Y.~Dong, S.~Yu, D.~Qiu, J.~Liu,
  M.~Bi, B.~Song, W.~Sun, J.~Zheng, B.~Zhao, Y.~Cao, J.~Yang, Y.~Cao, X.~Kong,
  J.~Yu, Y.~Xue, and Z.~Xie, \enquote{{NTIRE} 2022 spectral demosaicing
  challenge and data set,} {\protect\JournalTitle{Proceedings of the
  {IEEE}/{CVF} Conference on Computer Vision and Pattern Recognition}} pp.
  882--896 (2022).

\bibitem{lukac_color_2005}
R.~Lukac and K.~Plataniotis, \enquote{Color filter arrays: design and
  performance analysis,} {\protect\JournalTitle{{IEEE} Trans. Consumer
  Electron.}} \textbf{51}, 1260--1267 (2005).

\bibitem{hirakawa_spatio-spectral_2008}
K.~Hirakawa and P.~Wolfe, \enquote{Spatio-spectral color filter array design
  for optimal image recovery,} {\protect\JournalTitle{{IEEE} Trans. on Image
  Process.}} \textbf{17}, 1876--1890 (2008).

\bibitem{li_optimized_2017}
J.~Li, C.~Bai, Z.~Lin, and J.~Yu, \enquote{Optimized color filter arrays for
  sparse representation-based demosaicking,} {\protect\JournalTitle{{IEEE}
  Trans. on Image Process.}} \textbf{26}, 2381--2393 (2017).

\bibitem{miao_design_2006}
L.~Miao and H.~Qi, \enquote{The design and evaluation of a generic method for
  generating mosaicked multispectral filter arrays,}
  {\protect\JournalTitle{{IEEE} Trans. on Image Process.}} \textbf{15},
  2780--2791 (2006).

\bibitem{li_optimized_2018}
Y.~Li, A.~Majumder, H.~Zhang, and M.~Gopi, \enquote{Optimized multi-spectral
  filter array based imaging of natural scenes,}
  {\protect\JournalTitle{Sensors}} \textbf{18}, 1172 (2018).

\bibitem{lapray_multispectral_2014}
P.~J. Lapray, X.~Wang, J.~B. Thomas, and P.~Gouton, \enquote{Multispectral
  filter arrays: Recent advances and practical implementation,}
  {\protect\JournalTitle{Sensors (Switzerland)}} \textbf{14}, 21626--21659
  (2014).

\bibitem{saxe_advances_2018}
S.~Saxe, L.~Sun, V.~Smith, D.~Meysing, C.~Hsiung, A.~Houck, M.~Von~Gunten,
  C.~Hruska, D.~Martin, R.~Bradley, J.~Amoroso, M.~Klimek, and B.~Houck,
  \enquote{Advances in miniaturized spectral sensors,} in \emph{Next-Generation
  Spectroscopic Technologies {XI},}  M.~A. Druy, R.~A. Crocombe, S.~M. Barnett,
  L.~T. Profeta, and A.~K. Azad, eds. ({SPIE}, 2018), p.~10.

\bibitem{pichette_fast_2017}
J.~Pichette, W.~Charle, and A.~Lambrechts, \enquote{Fast and compact internal
  scanning {CMOS}-based hyperspectral camera: the snapscan,} in
  \emph{Conference proceedings of spie,}  vol. 10110 ({SPIE}, 2017), p.
  1011014. Conference Name: Society of Photo-Optical Instrumentation Engineers
  ({SPIE}) Conference Series {ADS} Bibcode: 2017SPIE10110E..14P.

\bibitem{lemmens_combination_2020}
S.~Lemmens, T.~Van~Craenendonck, J.~Van~Eijgen, L.~De~Groef, R.~Bruffaerts,
  D.~A. de~Jesus, W.~Charle, M.~Jayapala, G.~Sunaric-Mégevand, A.~Standaert,
  J.~Theunis, K.~Van~Keer, M.~Vandenbulcke, L.~Moons, R.~Vandenberghe,
  P.~De~Boever, and I.~Stalmans, \enquote{Combination of snapshot hyperspectral
  retinal imaging and optical coherence tomography to identify alzheimer’s
  disease patients,} {\protect\JournalTitle{Alz Res Therapy}} \textbf{12}, 144
  (2020).

\bibitem{cheng_-line_2018}
H.~Cheng, C.~Deng, Y.~Li, N.~Liao, W.~Yang, and X.~Bai, \enquote{An on-line
  color defect detection method for printed matter based on snapshot
  multispectral camera,} in \emph{Advanced Optical Imaging Technologies,}
  X.-C. Yuan, K.~Shi, and M.~G. Somekh, eds. ({SPIE}, 2018), p.~39.

\bibitem{hagen_review_2013}
N.~Hagen and M.~W. Kudenov, \enquote{Review of snapshot spectral imaging
  technologies,} {\protect\JournalTitle{Opt. Eng}} \textbf{52}, 090901 (2013).

\bibitem{fu_hyperspectral_2020}
H.~Fu, L.~Bian, X.~Cao, and J.~Zhang, \enquote{Hyperspectral imaging from a raw
  mosaic image with end-to-end learning,} {\protect\JournalTitle{Opt. Express}}
  \textbf{28}, 314 (2020).

\bibitem{arad_ntire_2022-1}
B.~Arad, R.~Timofte, R.~Yahel, N.~Morag, A.~Bernat, Y.~Cai, J.~Lin, Z.~Lin,
  H.~Wang, Y.~Zhang, H.~Pfister, L.~V. Gool, S.~Liu, Y.~Li, C.~Feng, L.~Lei,
  J.~Li, S.~Du, C.~Wu, Y.~Leng, R.~Song, M.~Zhang, C.~Song, S.~Zhao, Z.~Lang,
  W.~Wei, L.~Zhang, R.~Dian, T.~Shan, A.~Guo, C.~Feng, J.~Liu, M.~Agarla,
  S.~Bianco, M.~Buzzelli, L.~Celona, R.~Schettini, J.~He, Y.~Xiao, J.~Xiao,
  Q.~Yuan, J.~Li, L.~Zhang, T.~Kwon, D.~Ryu, H.~Bae, H.-H. Yang, H.-E. Chang,
  Z.-K. Huang, W.-T. Chen, S.-Y. Kuo, J.~Chen, H.~Li, and S.~Liu,
  \enquote{{NTIRE} 2022 spectral recovery challenge and data set,}
  {\protect\JournalTitle{Proceedings of the {IEEE}/{CVF} Conference on Computer
  Vision and Pattern Recognition}} pp. 863--881 (2022).

\bibitem{tao_compressive_2021}
C.~Tao, H.~Zhu, X.~Wang, S.~Zheng, Q.~Xie, C.~Wang, R.~Wu, and Z.~Zheng,
  \enquote{Compressive single-pixel hyperspectral imaging using {RGB} sensors,}
  {\protect\JournalTitle{Opt. Express}} \textbf{29}, 11207 (2021).

\bibitem{chakrabarti_learning_2016}
A.~Chakrabarti, \enquote{Learning sensor multiplexing design through
  back-propagation,} {\protect\JournalTitle{Advances in Neural Information
  Processing Systems}} pp. 3089--3097 (2016).

\bibitem{henz_deep_2018}
B.~Henz, E.~S. Gastal, and M.~M. Oliveira, \enquote{Deep joint design of color
  filter arrays and demosaicing,} {\protect\JournalTitle{Computer Graphics
  Forum}} \textbf{37}, 389--399 (2018).

\bibitem{jacome_d2uf_2022}
R.~Jacome, J.~Bacca, and H.~Arguello, \enquote{D2uf: Deep coded aperture design
  and unrolling algorithm for compressive spectral image fusion,}
  {\protect\JournalTitle{{IEEE} Journal of Selected Topics in Signal
  Processing}} pp. 1--11 (2022). Conference Name: {IEEE} Journal of Selected
  Topics in Signal Processing.

\bibitem{zhang_deeply_2021}
W.~Zhang, H.~Song, X.~He, L.~Huang, X.~Zhang, J.~Zheng, W.~Shen, X.~Hao, and
  X.~Liu, \enquote{Deeply learned broadband encoding stochastic hyperspectral
  imaging,} {\protect\JournalTitle{Light Sci Appl}} \textbf{10}, 108 (2021).
  Number: 1 Publisher: Nature Publishing Group.

\bibitem{song_deep-learned_2021}
H.~Song, Y.~Ma, Y.~Han, W.~Shen, W.~Zhang, Y.~Li, X.~Liu, Y.~Peng, and X.~Hao,
  \enquote{Deep-learned broadband encoding stochastic filters for computational
  spectroscopic instruments,} {\protect\JournalTitle{Advanced Theory and
  Simulations}} \textbf{4}, 2000299 (2021). \_eprint:
  https://onlinelibrary.wiley.com/doi/pdf/10.1002/adts.202000299.

\bibitem{li_jointly_2023}
K.~Li, D.~Dai, and L.~Van~Gool, \enquote{Jointly learning band selection and
  filter array design for hyperspectral imaging,} in \emph{2023 {IEEE}/{CVF}
  Winter Conference on Applications of Computer Vision ({WACV}),}  (2023), pp.
  6373--6383. {ISSN}: 2642-9381.

\bibitem{hore_image_2010}
A.~Horé and D.~Ziou, \enquote{Image quality metrics: {PSNR} vs. {SSIM},} in
  \emph{2010 20th International Conference on Pattern Recognition,}  (2010),
  pp. 2366--2369.

\bibitem{abadi_tensorflow_2016}
M.~Abadi, P.~Barham, J.~Chen, Z.~Chen, A.~Davis, J.~Dean, M.~Devin,
  S.~Ghemawat, G.~Irving, M.~Isard, M.~Kudlur, J.~Levenberg, R.~Monga,
  S.~Moore, D.~G. Murray, B.~Steiner, P.~Tucker, V.~Vasudevan, P.~Warden,
  M.~Wicke, Y.~Yu, and X.~Zheng, \enquote{{TensorFlow}: A system for
  large-scale machine learning,} in \emph{Proceedings of the 12th {USENIX}
  Symposium on Operating Systems Design and Implementation,}  (2016), p.~21.

\bibitem{kingma_adam_2017}
D.~P. Kingma and J.~Ba, \enquote{Adam: A method for stochastic optimization,}
  (2017).

\bibitem{esposito_-orbit_2019}
M.~Esposito, S.~S. Conticello, M.~Pastena, and B.~Carnicero~Dominguez,
  \enquote{In-orbit demonstration of artificial intelligence applied to
  hyperspectral and thermal sensing from space,} in \emph{Conference
  proceedings of spie,}  (2019), pp. 11--11.

\bibitem{simon_oxford_2013}
S.~H. Simon, \emph{The Oxford solid state basics} (Oxford University Press,
  2013), 1st ed. {OCLC}: ocn853504907.

\bibitem{acharya_index_1981}
B.~Acharya and M.~Gill, \enquote{On the index of gracefulness of a graph and
  the gracefulness of two-dimensional square lattice graphs,}
  {\protect\JournalTitle{Indian J. Math}} \textbf{23}, 14 (1981).

\bibitem{esposito_-orbit_2019-1}
M.~Esposito and A.~Zuccaro~Marchi, \enquote{In-orbit demonstration of the first
  hyperspectral imager for nanosatellites,} in \emph{International Conference
  on Space Optics — {ICSO} 2018,}  vol. 11180 N.~Karafolas, Z.~Sodnik, and
  B.~Cugny, eds. ({SPIE}, 2019), pp. 71--71. Issue: October 2018.

\bibitem{audebert_deep_2019}
N.~Audebert, B.~Le~Saux, and S.~Lefevre, \enquote{Deep learning for
  classification of hyperspectral data: A comparative review,}
  {\protect\JournalTitle{{IEEE} Geosci. Remote Sens. Mag.}} \textbf{7},
  159--173 (2019).

\bibitem{imani_overview_2020}
M.~Imani and H.~Ghassemian, \enquote{An overview on spectral and spatial
  information fusion for hyperspectral image classification: Current trends and
  challenges,} {\protect\JournalTitle{Information Fusion}} \textbf{59}, 59--83
  (2020).

\bibitem{rasti_feature_2020}
B.~Rasti, D.~Hong, R.~Hang, P.~Ghamisi, X.~Kang, J.~Chanussot, and J.~A.
  Benediktsson, \enquote{Feature extraction for hyperspectral imagery: The
  evolution from shallow to deep: Overview and toolbox,}
  {\protect\JournalTitle{{IEEE} Geosci. Remote Sens. Mag.}} \textbf{8}, 60--88
  (2020).

\bibitem{stockmans_end--end_2022}
T.~Stockmans, F.~Snik, M.~Smit, J.~Rietjens, M.~Esposito, C.~Van~Dijk, and
  C.~Keller, \enquote{End-to-end design framework for compressed on-chip
  pixel-wise spectro-polarimeters,} in \emph{{CubeSats} and {SmallSats} for
  Remote Sensing {VI},}  C.~D. Norton and S.~R. Babu, eds. ({SPIE}, 2022),
  p.~14.

\end{thebibliography}

\end{document}